\documentclass[aps,floatfix,showpacs]{revtex4}
\usepackage[latin1]{inputenc}
\usepackage{graphicx}
\usepackage{latexsym}
\usepackage{amstext}
\usepackage{amsxtra}
\usepackage{epsf}

\def\>{\rangle}
\def\<{\langle}

\def\psd{{{\hat\psi}^{\dagger}}}

\def\ps{\hat{\psi}}
\def\tet{\hat{\theta}}
\def\ro{\hat{\rho}}
\def\dro{\delta \hat{\rho}}
\def\hb{\hat{B}}
\def\hbd{\hat{B}^{\dagger}}
\def\hq{\hat{Q}}
\def\hP{\hat{P}}

\def\roo{\rho_0}

\def\kb{{\bf k}}
\def\rb{{\bf r}}
\def\rbp{{\bf r'}}

\begin{document}
\title{Extension of Bogoliubov theory to quasi-condensates}
\author{Christophe Mora}
\affiliation{Laboratoire de physique statistique, \'Ecole normale
sup\'erieure, 24 rue Lhomond,
75231 Paris Cedex 5, France.}
\author{Yvan Castin}
\affiliation{Laboratoire Kastler Brossel, \'Ecole normale
sup\'erieure, 24 rue Lhomond,
75231 Paris Cedex 5, France.}

\date{26 September 2002}

\begin{abstract}
We present an extension of the well-known Bogoliubov theory
 to treat low dimensional degenerate 
Bose gases in the limit of weak interactions and low density fluctuations. 
We use a density-phase representation and show that a precise definition of 
the phase operator requires a space discretisation in cells of size $l$.
We perform a systematic expansion of the Hamiltonian in terms of two small
parameters, the relative density fluctuations inside a cell
and the phase change over a cell.
The resulting macroscopic observables can be computed in one, two and 
three dimensions with no ultraviolet or infrared divergence. Furthermore 
this approach exactly matches Bogoliubov's approach when there is a true 
condensate. We give the resulting expressions for the equation of state of the 
gas, the ground state energy, the first order and second order correlations
functions of the field. Explicit calculations are done for homogeneous 
systems.
\end{abstract}


\maketitle

Recent progresses in the realization of low-dimensional Bose gases in the quantum degenerate
regime offer new perspectives for the comparison with theoretical treatments. In atomic Bose gases,  
low-dimensional systems are achieved by creating anisotropic trapping potentials.
Bose-Einstein condensates of reduced dimensionality, that is with the atomic motion
frozen in the harmonic oscillator ground state along one 
or two directions, have been produced \cite{ketterle,salomon}. 

Low-dimensional Bose gases have been the subject of early theoretical studies.
In the thermodynamical limit for spatially homogeneous systems, 
the Mermin-Wagner-Hohenberg theorem \cite{mermin,hohen} excludes
the formation of a Bose-Einstein condensate at finite temperature. This is physically
due to large phase fluctuations which restrict the coherence length of the bosonic
field to a finite value.
One expects however that strong enough repulsive interactions
between the particles strongly reduces the density fluctuations of
the gas in contrast to the non-interacting case \cite{yvan1,Iac}. In this context,
Popov introduced a long time ago the concept of {\it quasi-condensate}
\cite{popov}. This concept has been extended to trapped
gases \cite{1d,2d,jason}.  The recent observation of large phase
fluctuations for a degenerate Bose gas in a highly anisotropic cigar shaped
trap has brought a qualitative experimental confirmation of the theory in a quasi one-dimensional
geometry \cite{sengstock,aspect}.

It turns out that the theory of quasi-condensates has not reached
yet the maturity of the theory for condensates.
In the case of regular 3D Bose-Einstein condensation in the weakly interacting
regime, the Bogoliubov theory \cite{Bogo}, based on a systematic expansion in a small parameter,
gives a very precise description of the state of the gas. The intuitive
idea of the Bogoliubov theory is to use the existence of a single macroscopically
occupied mode $\phi_0$ of the field, the mode of the condensate. We recall
here the $U(1)$-symmetry preserving version of the theory \cite{gardiner,yvan2}.
One first
splits the bosonic field operator as $\ps = \phi_0 \hat{a}_0+ \delta \ps$,
where $\hat{a}_0$ annihilates a particle in the condensate mode and
$\delta \ps$ accounts for quantum and thermal fluctuations in the other modes.
Then one uses the assumption $| \delta \ps | \ll |\hat{a}_0|$ to solve
perturbatively the field equations of motion.
This approach is not suitable for a {\it quasi-condensate} as there is
no single macroscopically occupied field mode. Fortunately, in the case of
weak density fluctuations, 
the Bogoliubov idea can still be adapted in a quantum phase-density 
representation of the field operator. One writes the field operator $\ps$
as $\exp(i\hat{\theta})\hat{\rho}^{1/2}$ where $\hat{\theta}$ and $\hat{\rho}$
are position dependent operators giving the
phase and the density.  One then splits the operator giving the density
as $\rho_0 + \delta\hat{\rho}$, where $\rho_0$ is a c-number and
$\delta\hat{\rho}$ are fluctuations, and one uses the fact that $|\delta\hat{\rho}|\ll\rho_0$.
This idea has already been used in the literature \cite{shevchenko} but to our
knowledge without a precise definition of the phase operator, a well-known delicate point
of quantum field theory \cite{phase_operator,girardeau}. 
As a consequence of
the non-rigorous definition of the phase, divergences appear in the
theory \cite{shevchenko}: one has to introduce an arbitrary energy cut-off,
so that predictions in 1D at zero temperature are made within a logarithmic accuracy only,
and in 3D there is no full equivalence with the Bogoliubov theory.
Another approach based on the current-density operator rather than on the phase
operator is given by
Schwartz \cite{schwartz}: an expansion of the Hamiltonian in terms of
weak density and current fluctuations is performed
relating the correlation function of the field 
to the static structure factor. It is subject to the same divergence problem
in 2D and 3D
in the absence of energy cut-off if one calculates the structure factor
in the Bogoliubov approximation.

A possibility to circumvent these difficulties is to rely on the path
integral formulation of quantum field theory, which involves a functional
integral over a classical field, for which the phase is perfectly well defined.
This is the approach used by Popov, but with the introduction of 
an energy cut-off much smaller than the chemical potential of the gas,
so that the physics at length scales smaller than the healing
length is not accurately described.  
This functional integral approach has recently been revisited \cite{stoof1,stoof2}: it can 
lead to a cut-off
independent formalism for quasi-condensates. The predictions of \cite{stoof1,stoof2}
have the unsatisfactory feature of not
reproducing exactly the same results as the Bogoliubov theory for a 
3D condensate, but this has been corrected in an erratum \cite{erratum}.
 
In this paper, we propose a new improved Bogoliubov approach to treat {\it quasi-condensates} 
in the phase-density formalism for a weakly-interacting Bose gas.
This approach is based on a lattice model, that is with discrete spatial modes, which allows
to give a careful definition of the phase operator of the field and to introduce from
the start an energy cut-off.
It uses a systematic expansion in powers of the density fluctuations and of the spatial
phase gradient and leads
to simple expressions for the first order and the second order spatial correlation functions
of the bosonic field that do not depend on the energy cut-off and that exactly reproduces
in 3D the predictions of the Bogoliubov theory.
We also use this formalism to determine the equation of state of the gas
to the lowest non-vanishing order in the thermal and quantum excitations.

In section \ref{secnew}, we construct a discretized space model 
in order to define in a precise way the operators giving the phase and the density.
We give the physical implications  of the space discretization, restricting this  approach to 
highly degenerate and weakly interacting Bose systems. 
In section \ref{modes}, we derive a quadratic approximation to the Hamiltonian, that is
we derive approximate linear equations of motion for the density fluctuations and the phase
operators. We recover to the lowest order the Gross-Pitaevskii equation for the {\it quasi-condensate}
density and we recover the Bogoliubov spectrum for the excitations. 
We also push the expansion to the next order, by
producing a cubic correction to the quadratic Hamiltonian,
including the interaction between the {\it quasi-condensate} and the excitations. 
We show that inclusion of this correction is necessary
to get a consistent theory and to establish the full equivalence
between our approach and the number conserving Bogoliubov theory. 
In section \ref{application}, we present a few applications of our formalism:
we give general formulas for the equation of state and the ground state
of the gas, and for the first order and second order correlation functions
$g_1$ and $g_2$ of the field operator.
In section \ref{homogene}, we apply our formal results to the homogeneous Bose gas 
in various dimensions of space. This allows to derive simply the validity
condition of the method and to compare with existing results in the literature.

\section{Construction of a discrete phase-density representation}\label{secnew}
\subsection{Why discretize the real space ?}\label{subsec:why}
In previous studies of {\it quasi-condensates} the basic tools of the theory
are an operator 
\begin{equation}
\hat{\rho}(\rb)=\hat{\psi}^{\dagger}(\rb)
\hat{\psi}(\rb)
\end{equation}
giving the density in $\rb$ and an operator $\hat{\theta}(\rb)$
giving the phase of $\hat{\psi}(\rb)$, 
the field operator in $\rb$, the position $\rb$ being a continuous
variable \cite{griffin}. A small parameter of the theory characterizing the regime
of {\it quasi-condensates} is then that the density fluctuations, that is the fluctuations
of $\hat{\rho}(\rb)$, are small in relative values:
\begin{equation}
\mbox{Var}(\hat{\rho}(\rb))\equiv \langle \hat{\rho}(\rb)^2\rangle - \langle \hat{\rho}(\rb) \rangle^2
\ll \langle \hat{\rho}(\rb) \rangle^2.
\end{equation}
However one finds that the expectation value of $\hat{\rho}(\rb)^2$ is infinite
in every point with a non-vanishing mean density $\rho(\rb)=\langle \hat{\rho}(\rb)\rangle$:
\begin{equation}
\langle \hat{\rho}(\rb)^2\rangle = \delta({\bf 0}) \rho(\rb)
+\langle
\hat{\psi}^\dagger(\rb) \hat{\psi}^\dagger(\rb) \hat{\psi}(\rb) \hat{\psi}(\rb)
\rangle
\end{equation}
where $\delta({\bf 0})$, the value of the Dirac distribution at the origin,
is infinite, and the second term in the right hand side, giving the probability
density of finding two atoms in the same point of space, is finite in any
realistic model.  Mathematically 
this divergence is due to the use
of the bosonic commutation relations of the field operators $\hat{\psi}(\rb)$ 
and $\hat{\psi}^\dagger(\rb)$ in the same point of space to put the
atomic field product in normal order.

In order to have small, and therefore finite, density fluctuations, one is
forced to discretize the space, that is to collect the particles in little boxes
at the nodes of a spatial grid. Each little box has equal length $l$ along each dimension of space
and is parameterized by the position $\rb$ of its center.
The field operator $\hat{\psi}(\rb)$ has the effect of removing a particle
in the box at position $\rb$ and it now satisfies the bosonic commutation
relations:
\begin{equation}\label{commu1}
\lbrack \ps ( \rb ), \psd ( \rbp ) \rbrack = \frac{\delta_{ \rb , \rbp }}{  l^D}
\end{equation}
where $\delta_{ \rb , \rbp }$ is the discrete Kronecker delta and $D$ is the
dimension of space. The variance of the operator giving the density
is now finite:
\begin{eqnarray}
\mbox{Var}(\hat{\rho}(\rb)) &= &
\langle \hat{\psi}^\dagger(\rb) \hat{\psi}^\dagger(\rb) \hat{\psi}(\rb) \hat{\psi}(\rb)
\rangle-\rho^2(\rb) \nonumber \\
&+& \frac{\rho(\rb)}{l^D}.
\end{eqnarray}
In the validity domain of the theoretical approach of this paper, this variance will be 
much smaller than $\rho^2(\rb)$ because both the term on the first line of the right 
hand side and the term on the second line are small:
\begin{eqnarray}
\label{conda}
|\langle \hat{\psi}^\dagger(\rb) \hat{\psi}^\dagger(\rb) \hat{\psi}(\rb) \hat{\psi}(\rb)
\rangle-\rho^2(\rb)| \ll \rho^2(\rb) \\
\label{condb}
\rho(\rb) l^D \gg 1.
\end{eqnarray}

\subsection{The phase operator}
In usual continuous space theories an hermitian field phase operator $\hat{\theta}(\rb)$ is introduced
subject to the following commutation relation with the operator giving the density:
\begin{equation}
[\hat{\rho}(\rb),\hat{\theta}(\rbp)] = i\delta(\rb-\rbp).
\end{equation}
In our discrete model the desired commutation relation is modified into
\begin{equation}
[\hat{\rho}(\rb),\hat{\theta}(\rbp)] = i\frac{\delta_{\rb,\rbp}}{l^D}.
\label{commut}
\end{equation}

First we recall briefly that there exists actually no hermitian
operator $\hat{\theta}(\rb)$ satisfying strictly the above commutation relation.
 From the identity (\ref{commut}) one can indeed show that the operator
\begin{equation}
T(\alpha) \equiv e^{-i\alpha \hat{\theta}(\rb)},
\end{equation}
where $\alpha$ is any real number, is a translation operator for the density \cite{hint}:
\begin{equation}
\label{transla}
T(\alpha)^\dagger \hat{\rho}(\rb) T(\alpha) = \hat{\rho}(\rb) +\frac{\alpha}{l^D}.
\end{equation}
This identity contradicts two fundamental properties of $\hat{\rho}(\rb)$,
the positiveness and the discreteness of its spectrum \cite{continuous}.

We now proceed with the construction of a phase operator $\hat{\theta}(\rb)$ approximately
satisfying the commutation relation (\ref{commut}). The key ingredients allowing
such an approximate construction are 
(i) to be in the limit of a large occupation number of the considered
box of the lattice, and (ii) to construct the operator $e^{i\hat{\theta}}$ first, which,
according to (\ref{transla}) taken with $\alpha=-1$, simply reduces the number of
particles in the considered box by one.

In each spatial box we introduce the basis of Fock states $|n,\rb\rangle$
with exactly $n$ particles in the box. In this basis the field operators
have the following matrix elements:
\begin{eqnarray}
\nonumber \ps ( \rbp ) |n, \rb  \> & = & \frac{\delta_{ \rb , \rbp }}{l^{D/2}} \,
\sqrt{n} \, |n-1, \rb  \>, \\  \psd ( \rbp ) |n, \rb  \> & = &
\frac{\delta_{ \rb , \rbp }}{l^{D/2}} \, \sqrt{n+1} \, |n+1, \rb  \>
\end{eqnarray}
as a consequence of the commutation relation (\ref{commu1}).
The atomic density $\ro$ defined by $\ro ( \rb )  = \psd ( \rb ) \ps ( \rb )$
is diagonal in the Fock state basis:
\begin{equation}
\ro ( \rbp ) |n, \rb  \> = \delta_{ \rb , \rbp } \, \frac{n}{l^{D}}    \, |n, \rb \>.
\end{equation}
We then introduce the operator $\hat{A}$ defined by:
\begin{equation}
\ps ( \rb ) \equiv \hat{A} ( \rb ) \sqrt{\ro ( \rb ) }.
\end{equation}
In the Fock space $\hat{A} ( \rb )$ reduces by
one the number of particles $n$ in the box $\rb$: 
\begin{equation}
\hat{A} (\rbp) |n,\rb \> = \left( 1 - \delta_{n,0} \right) \delta_{\rb,\rbp} |n-1,\rb \>.
\end{equation}
Note that its action on the vacuum state of the box gives zero.
For each box $\rb$, the definition of $\hat{A}$ leads to the exact relations:
\begin{eqnarray}
\hat{A} \hat{A}^{\dagger} = I, \qquad \hat{A}^{\dagger} \hat{A} =
I - | 0 \> \< 0|  \qquad \textrm{and } \qquad \lbrack
\hat{A},\hat{A}^{\dagger} \rbrack =  | 0 \> \< 0|,
\end{eqnarray}
where $I$ is the identity operator and $ | 0 \> $ is the zero-particle state or
vacuum state in the box of center $\rb$. We find that the operator $\hat{A}$ is almost unitary,
{\it i.e.} it is effectively unitary for a physical state of the system with a negligible
probability of having an empty box. In what follows, we assume that this condition 
is satisfied, so that the projector $| 0 \> \< 0|$ can be neglected:
\begin{equation}
\label{cond_proba}
\mbox{occupation probability of}\ |n=0,\rb\rangle \ll 1.
\end{equation}

In this case, we write the approximately unitary operator $\hat{A}$ as:
\begin{equation}
\hat{A} ( \rb ) \simeq e^{i \tet ( \rb )} \qquad \textrm{with} \qquad
\tet^{\dagger} ( \rb ) \simeq \tet ( \rb )
\end{equation}
which amounts to writing the field operator as:
\begin{equation}\label{psi}
 \ps (\rb )  \simeq e^{i \tet (\rb  ) } \sqrt{\ro (\rb  )} .
\end{equation}
This should be understood as a formal writing, allowing for
example to recover the matrix elements of $\hat{A}$ and therefore
of the field operator $\hat{\psi}$ from the commutation relation (\ref{commut}).
We summarize below all the commutation relations of our
phase-density representation:
\begin{eqnarray}\label{commu2}
\lbrack \ro ( \rb ), \tet ( \rbp ) \rbrack \simeq  \frac{i \delta_{ \rb , \rbp }}{
l^D}  \qquad \lbrack \ro ( \rb ) ,  \ro ( \rbp ) \rbrack = 0  \qquad
\lbrack \tet ( \rb ), \tet ( \rbp ) \rbrack \simeq 0
\end{eqnarray}

We come back to the constraint (\ref{cond_proba}) at the basis
of the construction of $\exp(i\hat{\theta})$. 
A sufficient condition to have a low probability for a zero particle occupation in a box 
is obtained for a large mean number of particles in the box and with small
relative particle number fluctuations.  This is the regime that we wish to consider
in this paper.
We are therefore back to the discussion of the previous subsection
and to the conditions (\ref{conda},\ref{condb}) for weak density fluctuations.
In particular, the construction of the operator $\exp(i\hat{\theta})$ becomes
problematic in the limit $l\rightarrow 0$, that is in the continuous model.

\subsection{How to choose the grid spacing $l$}\label{subsec:how}

Working on a grid can be also seen as performing a coarse-grain average over all physical quantities on a scale $l$. This averaging suppresses the short wavelength modes (shorter than $l$) and thus introduces an energy cutoff:
\begin{equation}
E_{\rm cut} \simeq \frac{\hbar^2}{m l^2}.
\end{equation}
This cutoff is of no physical consequence if all characteristic energies ($\mu$, $k_{B} T$) 
are smaller, i.e., that $l$ is smaller than the corresponding characteristic lengths.
This leads, for instance,  to the following restrictions for $l$:
\begin{eqnarray}\label{restriction}
  l < \xi \qquad \textrm{and} \qquad l < \lambda
\end{eqnarray}
where 
\begin{equation}
\label{def_xi}
\xi = \frac{\hbar}{\sqrt{m \mu}}
\end{equation}
is the healing length,
and
\begin{equation}
\lambda = \sqrt{\frac{2 \pi \hbar^2}{m k_{B} T}}
\end{equation}
is the thermal de Broglie wavelength.
These two restrictions, combined with Eq. (\ref{condb}), impose:
\begin{eqnarray}\label{temp}
\rho \lambda^{D} \gg 1, \\ \label{inter} \rho \xi^{D} \gg 1.
\end{eqnarray}
These are conditions of validity for our discrete model.

The first one, Eq. (\ref{temp}), is the quantum degeneracy regime occurring at sufficiently low temperatures. The second restriction, Eq. (\ref{inter}), corresponds to 
the regime of weakly interacting systems. Its dependence on the density varies according
to the dimension of space.
In 1D and 3D, the mean field prediction for the chemical potential
is $\mu\simeq g\rho$ where $g$ is a constant characterizing the interaction potential,
the so-called coupling constant. In 1D,
Eq.(\ref{inter}) is the high density limit where a mean field theory is valid;
we recall that the small density limit $\rho \, \xi = \hbar \sqrt{\rho / m g} \ll 1$
corresponds to the strongly interacting (or strongly correlated)
Tonks gas regime.
In 3D, the effective coupling constant $g$ is
related to the $s$-wave scattering length $a$ of the interaction potential, $g=4\pi\hbar^2a/m$,
so that $\rho \, \xi^3 \propto 1/ \sqrt{\rho a^3} \gg 1$: one recovers 
the usual small gaseous parameter $\sqrt{\rho a^3}$.
In 2D, the chemical potential scales as $\hbar^2\rho/(m\ln(1/\rho a^2))$ where $a$ is the scattering
length of the 2D interaction potential  so that the condition $\rho \xi^2 \gg 1$ results
in a low density condition, $\ln(1/\rho a^2) \gg 1$.

\section{Perturbative treatment of a model Hamiltonian}\label{modes}

\subsection{Model Hamiltonian}

In our lattice model, we represent the binary interaction potential among the
particles by a discrete delta potential:
\begin{equation}\label{potential}
V(\rb_1-\rb_2) = \frac{g_0}{l^D} \delta_{\rb_1,\rb_2}
\end{equation}
where $g_0$ is the bare coupling constant. Note that $g_0$ in general differs from
the effective coupling constant $g$, and we shall come back to this point
in section \ref{etat}.
With this model potential, the grand canonical Hamiltonian is:
\begin{equation}\label{grandH}
H  = \sum_{\rb  } l^D \Big \lbrack -\frac{\hbar^2 }{2 m} \psd (\rb ) 
\Delta \ps (\rb) + (U(\rb) - \mu) \psd (\rb)  \ps (\rb) +
\frac{g_0}{2} \psd (\rb) \psd (\rb) \ps (\rb) \ps (\rb)
\Big \rbrack
\end{equation}
where $U( \rb )$ is an external trapping potential and
where the Laplacian is a symmetric operator coupling the different neighboring boxes:
\begin{equation}\label{def2}
\Delta  f(\rb) =  \sum_{j} \frac{f(\rb+l{\bf e_j}) +
f(\rb- l {\bf e_j}) - 2 f(\rb ) }{l^2}.
\end{equation}
The ${\bf e_j}$ are the unitary vectors and $j$ the different orthogonal space 
directions (for example, $j=x,y,z$ in 3D).
As usual we take periodic boundary conditions inside a rectangular box with lengths integer
multiple of $l$.

We now rewrite the Hamiltonian in the density-phase representation, that is in term
of the operators $\hat{\rho}$ and $\hat{\theta}$ giving the density and
the phase as defined in the previous section.
The contributions of the trapping potential and of the interaction potential
to the Hamiltonian are local in real space and therefore include the operator
$\hat{\rho}$ only:
\begin{equation}
H_{\rm pot} = \sum_{\rb} l^D \hat{\rho}(\rb)\left[U(\rb)-\mu+\frac{g_0}{2}
\left(\hat{\rho}(\rb)-\frac{1}{l^D}\right)\right] 
\label{pot}
\end{equation}
where we have used the bosonic commutation relation (\ref{commu1}) to exchange
one of the $\hat{\psi}^\dagger$ with $\hat{\psi}$ in the interaction term.
The kinetic energy term involves explicitly the phase operator:
\begin{equation}\label{hkin}
H_{\rm kin} = -\frac{\hbar^2}{2ml^2}\sum_{\rb} l^D\, \sum_j \sqrt{\hat{\rho}}
\left(e^{i(\hat{\theta}_{+j}-\hat{\theta})}\sqrt{\hat{\rho}_{+j}}
+e^{i(\hat{\theta}_{-j}-\hat{\theta})}\sqrt{\hat{\rho}_{-j}}-2\sqrt{\hat{\rho}}
\right)
\end{equation}
where we have introduced the notations 
$\tet_{\pm j} = \tet (\rb  \pm l   {\bf e_j})$ and
$\ro_{\pm j} = \ro (\rb  \pm l   {\bf e_j})$.
A remarkable property of this writing, to be used below, is that it involves only the difference
of two phase operators between two neighboring points of the lattice.

\subsection{Hamiltonian quadratisation and cubisation}\label{Hquadra}

We now expand the Hamiltonian to third order
in powers of two small parameters. 
As already discussed in subsection  \ref{subsec:why} the regime of quasi-condensates
that we are interested in corresponds to small relative fluctuations $\delta\hat{\rho}$
of the density. In the zeroth order approach totally neglecting 
the density fluctuations,
the density is set to a deterministic value $\rho_0$, as we shall see.
The second order expansion allows us to describe  the density fluctuations:
\begin{equation}
\label{split_rho}
\hat{\rho}(\rb) = \rho_0(\rb) +\delta\hat{\rho}(\rb).
\end{equation}
The third order expansion allows us to calculate the mean value of
$\delta\hat{\rho}(\rb)$.

The first small parameter of the systematic expansion used in
this paper is therefore given by
\begin{equation}
\epsilon_1 = \frac{|\delta\hat{\rho}|}{\rho_0} \ll 1.
\label{eps1}
\end{equation}
where $|\delta\hat{\rho}|$ is the typical value of the 
operator $\delta\hat{\rho}$ in the physical state of the system.
Mathematically this allows to expand $\sqrt{\hat\rho}$ as
\begin{equation}
\label{exp_rho}
\hat{\rho}^{1/2} = \rho_0^{1/2} + \frac{1}{2} \frac{\delta\hat{\rho}}{\rho_0^{1/2}}
-\frac{1}{8}\frac{\delta\hat{\rho}^2}{\rho_0^{3/2}}+
\frac{1}{16}\frac{\delta\hat{\rho}^3}{\rho_0^{5/2}}\ldots
\end{equation}
The second small parameter of the expansion is given by
\begin{equation}
\epsilon_2 = |l{\bf\nabla}\hat{\theta}| \ll 1.
\label{eps2}
\end{equation}
Here $\bf\nabla$ represents the gradient on the
lattice:
\begin{equation}\label{def_nabla}
{\bf\nabla} f(\rb  ) =  \sum_j \frac{f(\rb  + l   {\bf e_j}) - f(\rb  
- l   {\bf e_j}) }{2 l}{\bf e_j}
\end{equation}
where $f$ is an arbitrary function.
Physically the existence of the small parameter $\epsilon_2$ is reasonable: 
it is at the basis
of our hypothesis that the continuous quantum field problem can be well
approximated by a discrete lattice model, provided that $l$ is small enough,
see subsection \ref{subsec:how}.
Mathematically this second small parameter
allows to expand in (\ref{hkin}) the exponentials of the phase differences:
\begin{equation}
\label{exp_theta}
e^{i(\hat{\theta}_{+j}-\hat{\theta})} = 1+i(\hat{\theta}_{+j}-\hat{\theta})-\frac{1}{2}
(\hat{\theta}_{+j}-\hat{\theta})^2 \ldots
\end{equation}

 From the fact that the discretisation length $l$ is on the order of the
smallest of the two macroscopic length scales $\xi$ and $\lambda$,  see (\ref{restriction}),
it will be checked later that the parameters $\epsilon_1$ and $\epsilon_2$,
though of an apparently different physical origin, can be chosen
to be of the same order of magnitude,
\begin{equation}
\epsilon_1 \sim \epsilon_2 \sim \frac{1}{\sqrt{\rho_0 l^D}}
\end{equation}
and can therefore be treated mathematically as infinitesimals of the same order.
The mathematical details of the expansion 
\begin{equation}
H= H_0 + H_1 + H_2 + H_3 +\ldots
\end{equation}
are given in the appendix \ref{appen:expansion},
we present here only the results:
\begin{eqnarray}
\label{H0}
\nonumber H_0 & = &   \sum_{\rb} l^D  \Big \lbrack 
-\frac{\hbar^2}{2 m} \sqrt{\rho_0}\Delta\sqrt{\rho_0}
+\frac{g_0}{2} \rho_0^2 +( U(\rb) - \mu ) 
\rho_0  
\Big \rbrack \\ 
\label{H1}
\nonumber H_1 & = & \sum_{\rb} l^D 
\Big\lbrack 
- \frac{\hbar^2}{2 m \sqrt{\rho_0}} \Delta \sqrt{\rho_0}   + U(\rb) - \mu + g_0 \rho_0
\Big\rbrack \dro  \\  
\label{H2}
\nonumber H_2 & = &  E_2[\rho_0] + \sum_{\rb} l^D \Big \lbrack 
- \frac{\hbar^2}{2 m} \frac{\dro}{2 \sqrt{\rho_0}} \Delta
\left(\frac{\dro}{2 \sqrt{\rho_0}}\right) + 
\frac{\hbar^2 \dro^2}{8 m \rho_0^{3/2}} 
\Delta \sqrt{\rho_0} + \frac{g_0}{2} \dro^2 \\  & + &
\frac{\hbar^2}{2 m}  \sum_{j} 
\sqrt{\rho_0(\rb  ) \rho_0(\rb   + l  {\bf e_j})} 
\, \frac{(\tet(\rb+l{\bf e_j})-\tet(\rb))^2}{l^2} \Big \rbrack \\
\label{H3}
\nonumber H_3 &=& -\frac{g_0}{2}
\sum_{\rb}\delta\hat{\rho} 
+\frac{\hbar^2}{4ml^2}
\sum_{\rb,j} l^D\,
(\hat{\theta}_{+j}-\hat{\theta})
\left(
\frac{\rho_{0,+j}^{1/2}}{\rho_0^{1/2}}\,\delta\hat{\rho}+
\frac{\rho_0^{1/2}}{\rho_{0,+j}^{1/2}}\,\delta\hat{\rho}_{+j} \right)
(\hat{\theta}_{+j}-\hat{\theta})
 \\
\nonumber &+&
\frac{\hbar^2}{8m} \sum_{\rb} 
\frac{\delta\hat{\rho}}{\rho_0}
\left( \rho_0^{-1/2}\Delta\rho_0^{1/2}-\rho_0^{1/2}\Delta\rho_0^{-1/2} \right)
-\frac{\hbar^2}{16m}\sum_{\rb} l^D
\left[\frac{\delta\hat{\rho}^3}{\rho_0^{5/2}}\Delta\sqrt{\rho_0}
-\frac{\delta\hat{\rho}^2}{\rho_0^{3/2}}\Delta\left(\frac{\delta\hat{\rho}}{\sqrt{\rho_0}}\right)
\right].
\end{eqnarray}
The quantity $E_2$ in $H_2$ is a c-number functional of the density $\rho_0$, given
in the appendix \ref{appen:expansion}, which has therefore no contribution to the
dynamics of the quantum field.

\subsection{Iterative solution for the quadratic Hamiltonian}\label{itersol}
We now solve perturbatively, order by order, 
the Hamiltonian problems defined by $H_0$, $H_0+H_1$ and $H_0+H_1+H_2$.
To zeroth order in $\epsilon_{1,2}$, the Hamiltonian is a c-number. As the chemical potential
is fixed in our approach, $H_0$ is minimized for a density profile $\rho_0(\rb)$
such that $\sqrt{\rho_0}$ solves the discrete version of the Gross-Pitaevskii equation:
\begin{equation}
\label{fond}
\left[-\frac{\hbar^2}{2 m } \Delta + U(\rb) - \mu + g_0  \rho_0\right]
\sqrt{\rho_0} = 0.
\end{equation}
This density profile constitutes the zeroth order approximation to the
density $\rho$. It contains a number of particles that we call
$N_0$:
\begin{equation}
N_0 \equiv \sum_{\rb} l^D \rho_0(\rb).
\end{equation}
Note that $N_0$ coincides with the mean total number of particles $N$ only
to lowest order in the theory.  Equation (\ref{fond}) defines $\rho_0$ and
therefore $N_0$ as function of the chemical potential $\mu$. It will however
turn out to be more convenient to parameterize the theory in terms of $N_0$
rather than in terms of $\mu$. We will therefore consider
$\mu$ and $\rho_0$ as functions of $N_0$:
\begin{eqnarray}
\label{mu0}\mu &=& \mu_0(N_0) \\
\rho_0(\rb) &=& \rho_0(\rb;N_0).
\end{eqnarray}
$\mu_0$ is therefore the Gross-Pitaevskii prediction for the chemical
potential of a gas of $N_0$ particles.

For the choice of density profile (\ref{fond}), 
the first order correction  $H_1$ to the Hamiltonian
vanishes. We therefore have now to solve the Hamiltonian problem
defined by $H_2$, in order to determine the lowest order
approximation to the density fluctuation $\delta\hat{\rho}$ and the phase $\hat{\theta}$.
It is instructive to write the corresponding Heisenberg equations of motion,
which are linear (and therefore trivially solvable) since $H_2$ is quadratic.
As $\tet$ and $\dro$ are two canonically conjugate variables, the equations
of motion are:
\begin{equation}
\begin{array}{rclcl}
\hbar\partial_t \tet & \simeq &  
\displaystyle -\frac{1}{l^D}
\frac{\partial H_2}{\partial \dro (\rb  )} &=&
\displaystyle \frac{\hbar^2}{2m\sqrt{\rho_0}}\left[
\Delta\left(\frac{\delta\hat{\rho}}{2\sqrt{\rho_0}}\right)-
\delta\hat{\rho} \frac{\Delta\sqrt{\rho_0}}{2\rho_0}
\right] - g_0\delta\hat{\rho} \\
\hbar \partial_t \dro (\rb) &\simeq&
\displaystyle\frac{1}{l^D} \frac{\partial H_2}{\partial \tet (\rb  )}  &=&
\displaystyle
-\frac{\hbar^2}{m}\sqrt{\rho_0}\left[
\Delta\left(\hat{\theta}\sqrt{\rho_0}\right)
-\hat{\theta}\Delta\sqrt{\rho_0} \right]
\end{array}
\end{equation}
An important difference of these equations with the so-called quantum hydrodynamics
equations for $\delta\hat{\rho}$ and $\hat{\theta}$
is that our formalism keeps the so-called quantum pressure term 
for $\partial_t\hat{\theta}$
whereas it is usually neglected 
in the literature \cite{griffin}. This allows our treatment to have a cut-off
energy larger than $\mu$, whereas the usual treatment is restricted to energy
modes much below $\mu$.

Furthermore one can simplify these equations using the Gross-Pitaevskii equation
(\ref{fond}) to eliminate $\Delta\sqrt{\rho_0}$:
\begin{eqnarray}
\label{hydro_1}
\hbar\partial_t \tet &=&
-\frac{1}{2\sqrt{\rho_0}}
\left[-\frac{\hbar^2}{2m}\Delta + U +3 g_0 \rho_0 -\mu\right]
\left(\frac{\dro}{\sqrt{\rho_0}}\right)\\
\label{hydro_2}
\hbar \partial_t \dro (\rb) &=& 
2\sqrt{\rho_0}\left[-\frac{\hbar^2}{2m}\Delta + U + g_0 \rho_0 -\mu \right]
\left(\hat{\theta}\sqrt{\rho_0}\right).
\end{eqnarray}
This gives the idea of a very simple canonical transformation which,
remarkably,
maps our equations for a quasi-condensate (\ref{hydro_1},\ref{hydro_2}) 
into the equations for the Bogoliubov
modes of a condensate: the field
\begin{equation}
\label{define_B}
\hb=\frac{\dro}{2 \sqrt{\rho_0}} + i \sqrt{\rho_0} \, \tet  
\end{equation}
has bosonic commutation relations 
\begin{equation}
\label{comm_B}
[\hb(\rb),\hbd(\rbp)]=\frac{\delta_{\rb,\rbp}}{l^D}
\end{equation}
and it obeys the standard Bogoliubov equations
\begin{equation}\label{Bogo}
i \hbar \partial_t \begin{pmatrix} \hb \\ \hbd  \end{pmatrix} =
{\cal L}_{\rm GP}
\begin{pmatrix} \hb \\ \hbd  \end{pmatrix}  \equiv
\begin{pmatrix}  -\frac{\hbar^2}{2m}\Delta + U - \mu + 2 g_0 \rho_0  & g_0 \rho_0 \\ - 
g_0\rho_0 &
 - \left( -\frac{\hbar^2}{2m}\Delta + U - \mu + 2 g_0 \rho_0 \right) 
\end{pmatrix} 
\begin{pmatrix} \hb \\ \hbd  \end{pmatrix}.
\end{equation}
This mapping can be readily extended to the Hamiltonian $H_2$, which is expected
to be canonically equivalent to the Bogoliubov Hamiltonian: 
\begin{equation}
\label{h2=hb}
H_2 =  l^D \sum_{\rb}
\hat{B}^\dagger\left(-\frac{\hbar^2}{2m}\Delta +U+g_0\rho_0-\mu \right)\hat{B}
+g_0\rho_0\left[\hat{B}^\dagger \hat{B} + \frac{1}{2}\left(\hat{B}^2
+\hat{B}^{\dagger 2} \right)
\right].
\end{equation}
We have checked that the identity (\ref{h2=hb}) indeed holds by replacing
$\hat{B}$ by its expression (\ref{define_B}) in terms of $\delta\hat{\rho}$
and $\hat{\theta}$, and by using the value of the commutators (\ref{commu2})
and the fact that $\sqrt{\rho_0}$ solves the Gross-Pitaevskii equation.
Remarkably the energy functional $E_2[\rho_0]$ is exactly compensated 
by the contribution of the commutators.

This mapping therefore allows to reuse the standard diagonalisation of the 
Bogoliubov Hamiltonian.
We recall here briefly the procedure described in \cite{lewenstein,yvan2}.
One introduces the normal eigenmodes $(u_s,v_s)$ of the Bogoliubov operator ${\cal L}_{\rm GP}$
with an energy $\epsilon_s$, normalisable as 
\begin{equation}
\sum_{\rb} l^D\, \lbrack |u_s (\rb )|^2 - |v_s (\rb )|^2 \rbrack  =  1.
\end{equation}
Then $(v_s^*,u_s^*)$ is an eigenmode of ${\cal L}_{\rm GP}$ with the energy $-\epsilon_s$.
To form a complete family of vectors one has to further introduce the zero energy mode
of ${\cal L}_{\rm GP}$, given by $(\phi_0,-\phi_0)$, and the anomalous mode 
$(\phi_a,\phi_a)$ with
\begin{eqnarray}
\phi_0 = \sqrt{\rho_0 / N_0} \qquad   \textrm{and} \qquad
\phi_a =  \sqrt{N_0} \partial_{N_{0}} \sqrt{\rho_0}  .
\label{phi_a}
\end{eqnarray}
The corresponding normalization of the anomalous mode is such that the scalar product
of $\phi_0$ and $\phi_a$ is 1/2.
With these definitions, one introduces the components of $(\hb,\hbd)$ on
the zero energy mode, on the anomalous mode and on the regular $(u_s,v_s)$ modes:
\begin{equation}\label{BBc}
\begin{pmatrix} \hb \\ \hbd  \end{pmatrix} = -i\sqrt{N_0} \hq
\begin{pmatrix} \phi_0 \\ - \phi_0 \end{pmatrix} + \frac{\hP}{\sqrt{N_0}}
\begin{pmatrix} \phi_a \\  \phi_a \end{pmatrix} + \sum_s \quad \hat{b}_s
\begin{pmatrix}  u_s \\ v_s  \end{pmatrix} + \hat{b}_s^{\dagger}
\begin{pmatrix}  v_s^* \\ u_s^*  \end{pmatrix}.
\end{equation}
$\hat{Q}$ is a collective coordinate representing the quantum phase of the field 
and $\hat{P}$ is its conjugate momentum 
\begin{equation}
\label{commPQ}
[\hat{P},\hat{Q}] = -i.
\end{equation}
Physically $\hat{P}$ corresponds to fluctuations
in the total number of particles, as expected, and as
shown in more details later, see (\ref{sensP}).
The operators  $\hat{b}_s$ are bosonic annihilation operators with the usual
commutation relations $[\hat{b}_s,\hat{b}_{s'}]=\delta_{s,s'}$. They commute
with $\hat{P}$ and $\hat{Q}$.
The inverse formulas giving $\hat{b}_s,\hat{Q}$ and $\hat{P}$ in terms
of $\hat{B}$ can be found for example in \cite{yvan2}.
Equation (\ref{BBc}) results in the following modal expansion
for the density fluctuations and the phase operators:
\begin{equation}\label{quant}
\begin{array}{rlc}
\tet (\rb  ) & = &   \displaystyle \sum_s
\theta_s (\rb  ) \, \hat{b}_s + \theta_s^* (\rb  ) \,
\hat{b}_s^{\dagger}  - {\hq} \\[3mm] \dro (\rb  ) & = &
\displaystyle \sum_s \delta \rho_s (\rb  ) \hat{b}_s + \delta
\rho_s^* (\rb  ) \hat{b}_s^{\dagger} + \hP \,
\partial_{N_0}\rho_0
\end{array}
\end{equation}
where 
\begin{eqnarray}\label{def1}
\nonumber \theta_s (\rb  ) &=& \frac{u_s (\rb  ) - v_s (\rb 
)}{2 i \sqrt{\roo(\rb  ) }}, \\ \delta \rho_s (\rb  ) &=& \sqrt{\roo(\rb
 )}  \, ( u_s (\rb  ) + v_s (\rb  ) ).
\end{eqnarray}
By construction, this modal expansion, when inserted into the quadratic Hamiltonian
$H_2$, results in 
\begin{equation}
\label{Hnorm}
H = \sum_s \epsilon_s \hat{b}_s^{\dagger} \hat{b}_s +
\frac{1}{2} \hP^2\mu_0' + \tilde{E}_2[\rho_0],
\end{equation}
where $\mu_0'=d\mu_0/dN_0$.
This is the sum of uncoupled harmonic oscillators,  plus a massive free degree of freedom
corresponding to the unbound phase variable $\hat{Q}$. The effective mass of
the phase variable is given by $1/\mu_0'$.
The energy functional $\tilde{E}_2[\rho_0]$ will be calculated in
\ref{gse}, where it will be shown that it leads to exactly the same 
ground state energy as the number conserving Bogoliubov theory.
This shows that the Bogoliubov theory can be used to calculate the ground state
energy even for e.g. 1D quasi-condensates, a fact commonly used in the literature
\cite{lieb,gaudin} but which looks rather heuristic in the absence of justification.

\subsection{Effect of cubic Hamiltonian corrections on the density}\label{cubic_corr}
The physics contained in the cubic term $H_3$ of the Hamiltonian is very rich. It includes
interaction effects between the Bogoliubov modes of the previous section,
allowing a generalization to quasi-condensates of the theory of energy
shifts and Beliaev-Landau damping usually put forward for Bose-Einstein
condensates \cite{Vincent,Stringari_Pitaevskii,Shlyapnikov}.

We are more modest here. Our motivation to include the cubic corrections
is that the quadratic Hamiltonian $H_2$ brings actually no correction
to the zeroth order approximation $\rho_0$ to the mean density,
since the mean value of $\delta\hat{\rho}$ vanishes at the level of the second order
theory. This is highly non satisfactory as it brings some inconsistency 
in the calculation of an observable like $g_1$, the first order correlation function
of the field: to get a non-trivial prediction for $g_1$ one has to include
terms quadratic in the phase operator, which are second order in $\epsilon_2$,
which forces to also include second order corrections to the mean density,
as will become very explicit in section \ref{application}.

We therefore calculate the first order
correction to the equations of motion of $\delta\hat{\rho}$
and $\hat{\theta}$
due to the cubic Hamiltonian term $H_3$, and we take the average over the quantum state 
corresponding to the density operator at thermal equilibrium for the Hamiltonian
$H_2$. This gives source terms to add to the equations for the mean density
and phase derived from $H_2$. We leave the details of the calculations
for the appendix \ref{appen:mvt} and we give directly the result:
\begin{eqnarray}
\label{pour_rho}
\hbar\partial_t\langle\delta\hat{\rho}\rangle_3 &=&\rho_0^{1/2}
\left[-\frac{\hbar^2}{2m}\Delta+U+g_0\rho_0-\mu\right]\left(2\rho_0^{1/2}
\langle\hat{\theta}\rangle_3\right)\\
\label{pour_theta}
-2\hbar\sqrt{\rho_0}\partial_t\langle\hat{\theta}\rangle_3 &=&
\left[ -\frac{\hbar^2}{2m}\Delta + U +3g_0\rho_0 -\mu \right] 
\left(\frac{\langle\delta\hat{\rho}\rangle_3-\langle\hbd\hb\rangle_{2}}{\rho_0^{1/2}} \right)
\nonumber \\
&+& g_0\rho_0^{1/2}\langle 4\hbd\hb+\hb^2+B^{\dagger 2} \rangle_2-
2\langle \hat{P}^2\rangle_2\mu_0'\partial_{N_0}\sqrt{\rho_0}
\end{eqnarray}
where the thermal average $\langle\ldots\rangle_2$ 
is taken with the unperturbed Hamiltonian $H_2$ and
$\langle\ldots\rangle_3$ is taken with the perturbed Hamiltonian
$H_2+H_3$ to first order in $H_3$. The expectation value of the `kinetic
energy' of the unbound phase variable in (\ref{Hnorm}) is equal to $k_B T/2$
according to the equipartition theorem so that
\begin{equation}
\label{equipart}
\langle \hat{P}^2\rangle_2 = \frac{k_B T}{\mu_0'}.
\end{equation}

At equilibrium the expectation value 
of $\partial_t\delta\hat{\rho}$ and $\partial_t\hat{\theta}$ vanish. This fact
is obvious for $\partial_t\delta\hat{\rho}$; it is less obvious for $\partial_t\hat{\theta}$
because of the presence of the unbound variable $\hat{Q}$, we therefore produce
a proof of that in the appendix \ref{pas_de_courant}.
We therefore have to solve (\ref{pour_rho},\ref{pour_theta}) with the left-hand
side set to zero. The first equation (\ref{pour_rho}) imposes
that the mean value of $\hat{\theta}$ is position independent, a trivial result.
In the second equation, the operator acting on $\langle\delta\hat{\rho}\rangle_3$
is strictly positive so that it is invertible and (\ref{pour_theta})
determines the correction to the mean density in a unique way.

We now go through a sequence of transformations allowing us to get a
physical understanding of the value of $\langle\delta\hat{\rho}\rangle_3$.
The first step is to pull out the contribution of the `anomalous' terms
$\hat{P}$, $\hat{Q}$ in the modal expansion (\ref{BBc}):
\begin{equation}
\hat{B}(\rb) \equiv -i\sqrt{N_0} \hat{Q}\phi_0(\rb) + \frac{1}{\sqrt{N_0}}
\hat{P}\phi_a(\rb) + \hat{B}_{n}.
\end{equation}
We calculate the expectation values of (\ref{pour_theta}) involving the
operator $\hat{B}$, using the fact that all the crossed terms between the anomalous
part and the operators $b_s$ have a vanishing expectation value:
\begin{eqnarray}
\label{bdb}
\langle \hat{B}^\dagger \hat{B}\rangle_2 &=& 
\frac{\phi_a^2}{N_0} \langle\hat{P}^2\rangle_2 + 
N_0\phi_0^2 \langle\hat{Q}^2\rangle_2 + \langle \hat{B}^\dagger_n \hat{B}_n\rangle_2
-\phi_a \phi_0 \\
\label{bb}
\langle \hat{B}^2\rangle_2+\langle\hat{B}^{\dagger 2}\rangle_2 &=& 
2 \frac{\phi_a^2}{N_0} \langle\hat{P}^2\rangle_2 - 2 N_0\phi_0^2 \langle\hat{Q}^2\rangle_2
+\langle \hat{B}_n^2\rangle_2+\langle\hat{B}_n^{\dagger 2}\rangle_2.
\end{eqnarray}
The term $\phi_0\phi_a$ in (\ref{bdb}) comes from the non-commutation
of $\hat{P}$ and $\hat{Q}$, see (\ref{commPQ}). The contributions
of $\langle\hat{Q}^2\rangle_2$ in (\ref{bdb}) and (\ref{bb}),
when inserted into (\ref{pour_theta}), are shown to compensate exactly 
when one uses the fact that $\phi_0$ solves the Gross-Pitaevskii equation.
This was expected from the $U(1)$ symmetry of the Hamiltonian:
only differences of the phase operator in two points appear in the Hamiltonian, so that
$H$ does not depend on $\hat{Q}$ and the mean density 
does not depend on $\langle\hat{Q}^2\rangle_2$.

We therefore get an equation for $\langle\delta\hat{\rho}\rangle_3$
involving the expectation value of $\hat{P}^2$ as a source term, and which looks
rather involved: 
\begin{eqnarray}
0 &=& \left[ -\frac{\hbar^2}{2m}\Delta + U +3g_0\rho_0 -\mu \right]
\left(\frac{\langle\delta\hat{\rho}\rangle_3-\phi_a^2N_0^{-1}\langle\hat{P}^2\rangle_2
-(\langle\hbd_n\hb_n\rangle_{2}-\phi_0\phi_a)}{\rho_0^{1/2}} \right)  \nonumber \\
&+& g_0\rho_0^{1/2}\langle 4(\hbd_n\hb_n-\phi_0\phi_a)+\hb_n^2+B_n^{\dagger 2} \rangle_2+
\langle\hat{P}^2\rangle_2  
\left[6 g_0\rho_0^{1/2}\partial_{N_0}\sqrt{\rho_0}-2\mu_0'\right]
\partial_{N_0}\sqrt{\rho_0}
\label{complique}
\end{eqnarray}
Fortunately the underlying physics is very simple 
and allows to predict the effect of this source term on the mean density.
One first identifies the physical meaning of $\hat{P}$ in (\ref{quant}).
Using the well-known fact
that the eigenmodes of ${\cal L}_{\rm GP}$ are orthogonal 
for the modified scalar product of signature $(1,-1)$, one has \cite{yvan2}:
\begin{equation}
\label{sca0}
\langle\phi_0|u_s\rangle+\langle\phi_0|v_s\rangle
\equiv\sum_{\rb} l^D \phi_0(\rb)\left[u_s(\rb)+v_s(\rb)\right] = 0
\end{equation}
so that the sum of $\delta\rho_s$ over all spatial nodes vanishes.
As a consequence the operator $\hat{N}$ giving the total
number of particles in the gas is simply
\begin{equation}
\label{sensP}
\hat{N} \equiv \sum_{\rb} l^D \hat{\rho}(\rb) = N_0 + \hat{P}
\end{equation}
where we have used the identity
\begin{equation}
\sum_{\rb} l^D \partial_{N_0} \rho_0(\rb) = \frac{d}{dN_0}
\sum_{\rb} l^D\rho_0(\rb) = \frac{dN_0}{dN_0}=1.
\label{truc}
\end{equation}
The source terms involving $\hat{P}$ therefore correspond to fluctuations 
in the total number of particles in the gas, due to the fact that we 
consider the grand canonical ensemble. The effect of these grand canonical fluctuations
can be considered for the case of a pure quasi-condensate at the order of the present 
calculation so it is easy to calculate it directly. 
In the grand canonical ensemble the probability that the quasi-condensate 
has $n$ particles is
\begin{equation}
\Pi_n  \propto \exp\left[-\beta \left(E_0(n)-\mu n\right)\right]
\end{equation}
where $E_0(n)$ is the Gross-Pitaevskii energy for the density profile
$\rho_0(\rb;n)$:
\begin{equation}
\label{gpener}
E_0(n) = \sum_{\rb} l^D \left[
-\frac{\hbar^2}{2m}\sqrt{\rho_0(\rb;n)}
\Delta \sqrt{\rho_0(\rb;n)} + U(\rb)\rho_0(\rb;n)+ \frac{g_0}{2}\rho_0^2(\rb;n)
\right].
\end{equation}
The corresponding mean grand canonical density is
\begin{equation}
\rho_{\rm GC}(\rb) = \int dn\, \Pi_n \rho_0(\rb;n)
\end{equation}
where we treat $n$ as a continuous variable.
The zeroth order approximation $n=N_0$ for the number of particles in the
quasi-condensate is such that $E_0(n)-\mu n$ has a minimum:
\begin{equation}
\frac{d}{dn}[E_0(n)-\mu n] = \mu_0(n) - \mu =0\ \mbox{for}\ n=N_0
\end{equation}
as shown in (\ref{mu0}). The corresponding density profile is $\rho_0(\rb;N_0)$.
The next order correction to that is obtained by expanding the $n$-dependent
density profile to second order in $n-N_0$ and by averaging over $n$:
\begin{equation}
\delta\rho_{\rm GC} (\rb) = \langle (n-N_0)\rangle \partial_{N_0}
\rho_0(\rb;N_0) 
+\frac{1}{2} \langle(n-N_0)^2\rangle \partial_{N_0}^2 \rho_0(\rb;N_0).
\end{equation}
The second moment of $n-N_0$ is calculated to lowest non-vanishing order
by a Gaussian approximation to $\Pi_n$:
\begin{equation}
E_0(n) - \mu n \simeq E_0(N_0)-\mu N_0 +\frac{1}{2} \frac{d^2E_0}{dN_0^2}(n-N_0)^2
=\mbox{const}+\frac{1}{2}\mu_0' (n-N_0)^2.
\end{equation}
This leads to 
\begin{equation}
\langle (n-N_0)^2\rangle_{\rm Gauss} = \frac{k_B T}{\mu_0'} = 
\langle \hat{P}^2\rangle_2.
\end{equation}
More care has to be taken in the calculation of the mean of $n-N_0$: the Gaussian
approximation to $\Pi_n$ gives a vanishing contribution, so that the cubic distortion
to it has to be included:
\begin{eqnarray}
E_0(n) - \mu n &\simeq& E_0(N_0)-\mu N_0 + \frac{1}{2}\mu_0' (n-N_0)^2
+\frac{1}{6}\mu_0'' (n-N_0)^3  \\
\Pi_n &\propto& \exp\left[-\frac{1}{2}\beta \mu_0'(n-N_0)^2\right]\left[
1-\frac{1}{6}\beta\mu_0'' (n-N_0)^3\right].
\end{eqnarray}
We then get a non-vanishing mean value for $n-N_0$:
\begin{equation}
\langle (n-N_0)\rangle_{\rm distor}
 = -\frac{1}{6} \beta\mu_0'' \langle (n-N_0)^4\rangle_{\rm Gauss}
=-\frac{1}{2} \beta\mu_0'' \left(\langle \hat{P}^2\rangle_2\right)^2.
\end{equation}
We have therefore predicted in a very simple way the correction to the mean
density due to grand canonical fluctuations:
\begin{equation}
\label{drho_gc}
\delta\rho_{\rm GC}(\rb) =  \frac{1}{2}\langle\hat{P}^2\rangle_2
\left[\partial_{N_0}^2 \rho_0(\rb;N_0)-\frac{\mu_0''}{\mu_0'}
\partial_{N_0}\rho_0(\rb;N_0)\right].
\end{equation}

How does this compare to the general formalism (\ref{complique})? We need to
obtain a partial differential equation for $\delta\rho_{\rm GC}$. We just
take the second order derivative of the Gross-Pitaevskii
equation (\ref{fond}) with respect to $N_0$ and we
replace $\rho_0$ by $\sqrt{\rho_0}^2$ in the resulting equation and in
(\ref{drho_gc}). This leads to the remarkable identity
\begin{equation}
\left[ -\frac{\hbar^2}{2m}\Delta + U +3g_0\rho_0 -\mu \right]
\left(\frac{\delta\rho_{\rm GC}-N_0^{-1}\phi_a^2\langle\hat{P}^2\rangle_2} 
{\rho_0^{1/2}} \right)=
-\langle\hat{P}^2\rangle_2
\left[6 g_0\rho_0^{1/2}\partial_{N_0}\sqrt{\rho_0}-2\mu_0'\right]
\partial_{N_0}\sqrt{\rho_0}.
\end{equation}
The right-hand side of this identity coincides with the source term
of (\ref{complique}) involving $\hat{P}$! We have therefore successfully identified
$\delta\rho_{\rm GC}$ as a piece of $\langle\delta\hat{\rho}\rangle_3$
and we are left with the simpler equation
\begin{eqnarray}
0 &=& \left[ -\frac{\hbar^2}{2m}\Delta + U +3g_0\rho_0 -\mu \right]
\left(\frac{\langle\delta\hat{\rho}\rangle_3-\delta\rho_{\rm GC}
-(\langle\hbd_n\hb_n\rangle_{2}-\phi_0\phi_a)}{\rho_0^{1/2}} \right)  \nonumber \\
&+& g_0\rho_0^{1/2}\langle 4(\hbd_n\hb_n-\phi_0\phi_a)+\hb_n^2+B_n^{\dagger 2} \rangle_2.
\label{plussimple}
\end{eqnarray}

We are not totally satisfied yet since the operator $\hat{B}_n$ does not obey bosonic
commutation relations when the system is not spatially homogeneous, in particular
the field $\hat{B}_n$ does not commute with
itself when taken in two different points:
\begin{eqnarray}
[\hat{B}_n(\rb),\hat{B}_n(\rb')] &=& \phi_a(\rb)\phi_0(\rb')-\phi_a(\rb')\phi_0(\rb). \\
{[\hat{B}_n(\rb),\hat{B}_n^{\dagger}(\rb')]} &=&
\frac{1}{l^D} \delta_{\rb,\rb'}-\phi_0(\rb)\phi_a(\rb')-\phi_0(\rb')\phi_a(\rb).
\end{eqnarray}
To circumvent this difficulty we split the field $\hat{B}_n$ in its component
along the quasi-condensate mode $\phi_0$ and its orthogonal component:
\begin{equation}
\label{def_lambda}
\hat{B}_n(\rb) = \hat{\alpha} \phi_0(\rb) + \hat{\Lambda}(\rb).
\end{equation}
The Bogoliubov functions $u_s(\rb)$ and $v_s(\rb)$ can be chosen here to be real. 
The operator $\hat{\alpha}$ can be then written as
\begin{equation}
\hat{\alpha}=\sum_s \langle\phi_0|u_s\rangle \left(\hat{b}_s-\hat{b}_s^\dagger\right)
\end{equation}
where we have used the property (\ref{sca0}). 
This clearly shows that the operator $\hat{\alpha}$
is anti-Hermitian:
\begin{equation}
\hat{\alpha}^\dagger = -\hat{\alpha}.
\end{equation}
The field $\hat{\Lambda}$ has the following expansion on the $\hat{b}_s$:
\begin{equation}
\label{devLam}
\hat{\Lambda}(\rb) = \sum_s u_{s\perp}(\rb) \hat{b}_s 
+ v_{s\perp}(\rb)\hat{b}_s^\dagger
\end{equation}
where the index $\perp$ indicates projection orthogonally to $\phi_0$.
This field has now the desired bosonic commutation relations
\begin{eqnarray}
[\hat{\Lambda}(\rb),\hat{\Lambda}({\bf r}')] &=& 0 \\
{[\hat{\Lambda}(\rb),\hat{\Lambda}^{\dagger}({\bf r}')]} &=&
\frac{1}{l^D} \delta_{\rb,{\bf r}'}-\phi_0(\rb)\phi_0({\bf r}').
\label{commLam}
\end{eqnarray}
Note that $\hat{\alpha}$ does not commute with $\hat{\Lambda}$:
\begin{equation}
\label{bl}
[\hat{\alpha},\hat{\Lambda}(\rb)] = \frac{1}{2}\phi_0(\rb)-\phi_a(\rb).
\end{equation}

We insert the splitting of $\hat{B}_n$ in (\ref{plussimple}).
The terms quadratic in $\hat{\alpha}$ cancel exactly, in the same
way the terms in $\hat{Q}^2$ canceled. The terms linear in $\hat{\alpha}$
can all be expressed in terms of the expectation value
of an anticommutator, $\langle\{\hat{\alpha},\hat{\Lambda}\}\rangle_2$ using
the commutation relation (\ref{bl}) and the fact that $\langle \hat{\alpha} \hat{\Lambda}
\rangle_2$ is a real quantity. Furthermore, using the techniques
of Appendix E of \cite{cartago}, as shown in the appendix \ref{appen:cartago},
one obtains a simple partial differential equation for the anticommutator:
\begin{equation}
\label{elim_alpha}
\left[-\frac{\hbar^2}{2m}\Delta +U+g_0\rho_0-\mu \right]
\langle\{\hat{\alpha},\hat{\Lambda}(\rb)\}\rangle_2=
-\sum_{\rbp} l^D\, g_0\rho_0(\rbp)\phi_0(\rbp)\langle
\{\hat{\Lambda}(\rbp)+\hat{\Lambda}^\dagger(\rbp),\hat{\Lambda}(\rb) \}
\rangle_2.
\end{equation}
Remarkably this allows to eliminate completely the operator $\hat{\alpha}$
in (\ref{plussimple})! We finally get an equation for the correction to the
mean density involving the operator $\hat{\Lambda}$ only:
\begin{equation}
0=\left[ -\frac{\hbar^2}{2m}\Delta + U +3g_0\rho_0 -\mu \right]
\left(\frac{\langle\delta\hat{\rho}\rangle_3-\delta\rho_{\rm GC}
-\langle\hat{\Lambda}^\dagger\hat{\Lambda}\rangle_{2}}{\phi_0} \right)  
+S(\rb)
\label{ouf}
\end{equation}
where we have introduced the source term
\begin{equation}
\label{source}
S(\rb)\equiv
g_0N_0\phi_0(\rb)\langle 
4\hat{\Lambda}^\dagger(\rb)\hat{\Lambda}(\rb)+\hat{\Lambda}^2(\rb)
+\hat{\Lambda}^{\dagger 2}(\rb) -\phi_0^2(\rb)\rangle_2
-\sum_{\rbp} l^D\, g_0\rho_0(\rbp)\phi_0(\rbp)\langle
\{\hat{\Lambda}(\rbp)+\hat{\Lambda}^\dagger(\rbp),\hat{\Lambda}(\rb) \}
\rangle_2.
\end{equation}

It will reveal convenient to introduce the function 
$\chi(\rb)$ defined in a unique way by
\begin{equation}
\label{for_chi}
0= \left[ -\frac{\hbar^2}{2m}\Delta + U +3g_0\rho_0 -\mu \right]\chi(\rb)
+\frac{1}{2}S(\rb).
\end{equation}
We then obtain the following final expression for the correction
to the mean density due to the cubic Hamiltonian terms $H_3$:
\begin{equation}
\label{resultat}
\langle\delta\hat{\rho}\rangle_3(\rb) = \delta\rho_{\rm GC}(\rb)
+ 2 \phi_0(\rb) \chi(\rb)
+ \langle \hat{\Lambda}^+(\rb)\hat{\Lambda}(\rb)\rangle_2.
\end{equation}

In the particular case where the gas is Bose condensed,
our general theory for quasi-condensates also applies, of course. 
One then expects that the result (\ref{resultat}) has already
been obtained for the condensate and can be given a clear
physical interpretation. This expectation is totally justified:
as shown in the appendix \ref{appen:CD}, the component of
$\chi(\rb)/N_0$ orthogonal to $\phi_0$ is the correction given in \cite{yvan2}
to the Gross-Pitaevskii condensate wavefunction $\phi_0$ 
due to the interaction with
the non-condensed particles; the component of $\chi$ along $\phi_0$
describes the condensate depletion, and 
$\langle \hat{\Lambda}^+\hat{\Lambda}\rangle_2$ is the mean density
of non-condensed particles.

\section{Applications of the formalism: general formulas}\label{application}

\subsection{Equation of state}\label{etat}

What is referred to as the {\it equation of state} of the gas is the expression 
of the chemical potential as function of the mean total number of particles $N$ and
the temperature $T$. It is useful in particular to predict properties
of an inhomogeneous gas within the local density approximation.

We therefore have now to calculate $\mu$ for the quasi-condensate.
This is equivalent to a calculation of $N_0$ as $\mu$ and $N_0$ are by
definition related through (\ref{mu0}). 
To lowest order of the theory 
one assumes a pure quasi-condensate with a density profile $\rho(\rb)
=\rho_0(\rb)$, where $\sqrt{\rho_0}$ solves the Gross-Pitaevskii equation
(\ref{fond}). One therefore gets $N=N_0$ so that $\mu=\mu_0(N)$.

The first non-vanishing correction to the density profile is given by
(\ref{resultat}). By integrating (\ref{resultat}) over space
we get the corresponding correction for the mean total number of particles:
\begin{eqnarray}
N &\equiv& N_0 +\delta N \\ \label{deltaN}
\delta N &\simeq& \delta N_{\rm GC} + l^D \sum_{\rb} 2\phi_0(\rb)\chi(\rb)
+l^D\sum_{\rb} \langle \hat{\Lambda}^\dagger(\rb)\hat{\Lambda}(\rb)\rangle_2.
\end{eqnarray}
The contribution to $\delta N$ due to our use of the grand canonical ensemble
can be calculated exactly from a spatial integration of (\ref{drho_gc}),
using the same technique as in (\ref{truc}):
\begin{equation}
\delta N_{\rm GC} = -k_B T \frac{\mu_0''}{2{\mu_0'}^2}.
\end{equation}
The contribution of the term involving $\chi$ can also be made explicit
by multiplication of (\ref{for_chi}) by $\phi_a(\rb)$ defined
in (\ref{phi_a}) and by spatial integration.
The function $\phi_a(\rb)$ is indeed known \cite{lewenstein} to solve the partial
differential equation
\begin{equation}
\label{de_maciek}
\left[ -\frac{\hbar^2}{2m}\Delta + U +3g_0\rho_0 -\mu \right]\phi_a=
N_0\mu_0' \phi_0(\rb)
\end{equation}
which can be checked easily, just by taking the derivative of
(\ref{fond}) with respect to $N_0$. This leads to
\begin{equation}
l^D \sum_{\rb} 2\phi_0(\rb)\chi(\rb) = -\frac{1}{N_0\mu_0'}
\sum_{\rb} l^D \phi_a(\rb) S(\rb)
\end{equation}
where the source term $S$ is known explicitly, see (\ref{source}).
We just have now to replace $N_0$ by $N-\delta N$ in (\ref{mu0}) 
and expand to first order in $\delta N$:
\begin{equation}
\mu = \mu_0(N-\delta N) \simeq \mu_0(N) -\delta N \mu_0'(N_0).
\end{equation}
We obtain the following expression
for $\mu$:
\begin{equation}
\mu \simeq \mu_0(N) + k_B T \frac{\mu_0''}{2\mu_0'}
- \mu_0'(N_0) \sum_{\rb}l^D \langle 
\hat{\Lambda}^\dagger(\rb)\hat{\Lambda}(\rb)\rangle_2+ 
\frac{1}{N_0} \sum_{\rb} l^D \phi_a(\rb) S(\rb).
\end{equation}
Equivalently we can replace the source term by its explicit expression to get
\begin{eqnarray}
\nonumber
\mu \simeq \mu_0(N) + k_B T \frac{\mu_0''}{2\mu_0'}
-\mu_0'(N_0) \left(\frac{1}{2}+\sum_{\rb}l^D \langle
\hat{\Lambda}^\dagger(\rb)\hat{\Lambda}(\rb)\rangle_2\right) 
+\sum_{\rb} l^D g_0 \left(\partial_{N_0}\rho_0 (\rb)\right)
\left(2\langle\hat{\Lambda}^\dagger\hat{\Lambda}\rangle_2+
\mbox{Re}\langle\hat{\Lambda}^2\rangle_2
\right)\\
-\sum_{\rb}l^D g_0 \phi_0^3(\rb)\langle
\{\hat{\Lambda}(\rb)+\hat{\Lambda}^\dagger(\rb),\hat{\gamma}\}\rangle_2
\label{muC}
\end{eqnarray}
where we have introduced the operator
\begin{equation}
\hat{\gamma} = \sum_{\rb} l^D \phi_a(\rb)\hat{\Lambda}(\rb)
\end{equation}
and we have used the identity
\begin{equation}
\mu_0' = \sum_{\rb} l^D g_0\phi_0^2(\rb)\partial_{N_0}\rho_0(\rb;N_0)
\end{equation}
obtained by performing the scalar product of both sides of (\ref{de_maciek})
with $\phi_0$.
Application to spatially homogeneous systems will be given in section
\ref{homogene}; in this case
both the operator $\hat{\gamma}$ and $\mu_0''$ vanish.

\subsection{Ground state energy}\label{gse}
We now show that the ground state energy of a quasi-condensate 
can be calculated with exactly the same Bogoliubov formula as
for the ground state energy of a condensate. 

We have to determine the ground state energy of $H_2$. We write it
as the expectation value of (\ref{h2=hb}) at zero temperature,
that is here in the vacuum of the $\hat{b}_s$
and of $\hat{P}$:
\begin{equation}
\mbox{ground}(H_2) = 
l^D \sum_{\rb}
\langle \hat{B}^\dagger 
\left(-\frac{\hbar^2}{2m}\Delta +U+g_0\rho_0-\mu \right)\hat{B}\rangle_2
+g_0\rho_0\langle \left[\hat{B}^\dagger \hat{B} + \frac{1}{2}\left(\hat{B}^2
+\hat{B}^{\dagger 2} \right) \right]\rangle_2.
\label{h2gr}
\end{equation}
We reproduce the transformation of subsection \ref{cubic_corr}.
We split $\hat{B}$ in an anomalous part involving $\hat{P}$, $\hat{Q}$,
plus the contribution of the antihermitian operator $\hat{\alpha}$
and of $\hat{\Lambda}$, the orthogonal component of the normal part.
In the first expectation value
of the right-hand side of (\ref{h2gr}) the operators 
$\hat{Q}$ and $\hat{\alpha}$ disappear as they 
come with the factor $\phi_0(\rb)$ in $\hat{B}$ and 
$\phi_0$ solves the Gross-Pitaevskii equation
(\ref{fond}). The expectation value of $\hat{P}^2$ in the ground state
of $H_2$ also vanishes, so that
\begin{equation}
\langle \hat{B}^\dagger
\left(-\frac{\hbar^2}{2m}\Delta +U+g_0\rho_0-\mu \right)\hat{B}\rangle_2=
\langle \hat{\Lambda}^\dagger
\left(-\frac{\hbar^2}{2m}\Delta +U+g_0\rho_0-\mu \right)\hat{\Lambda}\rangle_2.
\end{equation}
The same transformation is applied to the last expectation value in
(\ref{h2gr}). Remarkably the terms involving $\hat{\alpha}$ exactly cancel
when one uses the relations 
(\ref{bdb}), (\ref{bb}), (\ref{bl}) and (\ref{Areel}). This leads
to
\begin{equation}
\langle \left[\hat{B}^\dagger \hat{B} + \frac{1}{2}\left(\hat{B}^2
+\hat{B}^{\dagger 2} \right) \right]\rangle_2
=-\frac{1}{2}\phi_0^2 + 
\langle \left[\hat{\Lambda}^\dagger \hat{\Lambda} + \frac{1}{2}\left(\hat{\Lambda}^2
+\hat{\Lambda}^{\dagger 2} \right) \right]\rangle_2.
\end{equation}
The expectation values involving $\hat{\Lambda}$ are readily calculated
from the modal expansion (\ref{devLam}):
\begin{equation}
\mbox{ground}(H_2) = -\frac{1}{2}\sum_{\rb}l^D g_0\rho_0\phi_0^2
+\sum_s \langle v_{s\perp}|
\left[\left(
-\frac{\hbar^2}{2m}\Delta +U+2g_0\rho_0-\mu 
\right)|v_{s\perp}\rangle + g_0\rho_0 |u_{s\perp}\rangle
\right].
\end{equation}
As $(u_s,v_s)$ is an eigenvector of ${\cal L}_{\rm GP}$, $(u_{s\perp},
v_{s\perp})$ is an eigenvector of the operator ${\cal L}$ defined
in \cite{yvan2} and this expression
can be further simplified  to
\begin{equation}
\label{gh2}
\mbox{ground}(H_2) = 
-\frac{1}{2}\sum_{\rb}l^D g_0\rho_0\phi_0^2
-\sum_s \epsilon_s \langle v_{s\perp}|v_{s\perp}\rangle.
\end{equation}

The last step is to include the contribution of $H_0$ and to remove
the $-\mu\hat{N}$ term from the grand canonical Hamiltonian.
The ground state energy of the canonical Hamiltonian for $N$ particles
is therefore
\begin{equation}
E_{\rm ground}(N) \simeq \mu N + E_0(N_0) -\mu N_0 +\mbox{ground}(H_2)
\end{equation}
where $E_0$ is the Gross-Pitaevskii energy (\ref{gpener}).
As we did in subsection \ref{etat} we replace $N_0$ by $N-\delta N$, where
$\delta N$ is calculated from $H_3$, and we expand $E_0(N-\delta N)$ 
to first order in $\delta N$:
\begin{equation}
\mu N + E_0(N_0) -\mu N_0 \simeq E_0(N) -\delta N (\mu_0(N)-\mu) \simeq E_0(N).
\end{equation}
We recall that by definition $\mu=\mu_0(N_0)$.
The first term in the right hand side of (\ref{gh2}) amounts to performing a small change
in the Gross-Pitaevskii energy functional, expressing the fact that a given particle
interacts in the gas with $N-1$ particles so that the mean field term should
be proportional to $N-1$ rather than to $N$. The final expression for
the ground state energy is:
\begin{equation}
E_{\rm ground}(N) \simeq 
N\sum_{\rb} l^D \left[
-\frac{\hbar^2}{2m}\phi_0(\rb;N)
\Delta \phi_0(\rb;N) + U(\rb)\phi_0^2(\rb;N)+ \frac{1}{2}g_0(N-1)\phi_0^2(\rb;N)
\right] - \sum_s \epsilon_s \langle v_{s\perp}|v_{s\perp}\rangle.
\end{equation}
This exactly coincides with the Bogoliubov result, see e.g. equation (71)
of \cite{yvan2}.

\subsection{Second order correlation function}\label{subsecg2}

The second order correlation function of the atomic field is 
defined as:
\begin{equation}
g_2 (\rb) \equiv \< \psd(\rb) \psd({\bf 0})  \ps({\bf 0}) \ps(\rb) \>
\end{equation}
where we have taken for simplicity one of the two points as the origin
of the coordinates.
To calculate $g_2$ with the formalism of this paper we have to express
$g_2$ in terms of the operator $\hat{\rho}$ giving the density. 
This is achieved thanks to the
commutation relation (\ref{commu1}) of the bosonic field $\hat{\psi}$:
\begin{equation}
g_2 (\rb) = \langle \hat{\rho}(\rb)\hat{\rho}({\bf 0})\rangle -
\frac{\delta_{\rb,{\bf 0}}}{l^D}  \langle \hat{\rho}({\bf 0})\rangle.
\end{equation}
We then insert the splitting (\ref{split_rho}) of $\hat{\rho}$ in terms
of the quasi-condensate density $\rho_0$ and the fluctuations $\delta\hat{\rho}$:
\begin{equation}
g_2 (\rb) = \rho_0(\rb)\rho_0({\bf 0}) + 
\rho_0({\bf 0})\langle\delta\hat{\rho}(\rb)\rangle +
\rho_0(\rb)\langle\delta\hat{\rho}({\bf 0})\rangle
+\langle \delta\hat{\rho}(\rb) \delta\hat{\rho}({\bf 0})\rangle
-\frac{\delta_{\rb,{\bf 0}}}{l^D} 
\left[\rho_0({\bf 0})+\langle\delta\hat{\rho}({\bf 0})\rangle
\right].
\label{g2_long}
\end{equation}

This expression of $g_2$ is still exact. We now perform approximations consistent
with an expansion of $g_2$ up to second order in the small parameters $\epsilon_{1,2}$.
The expectation value of the term quadratic in $\delta\hat{\rho}$ is calculated 
within the thermal equilibrium for the quadratic Hamiltonian $H_2$. The expectation
value of $\delta\hat{\rho}$ is evaluated in subsection \ref{cubic_corr} by inclusion
of the cubic perturbation $H_3$. The contribution of
$\delta\hat{\rho}$ in the last term of (\ref{g2_long}) is
negligible as it is $\epsilon_1^4$ times smaller than the leading
term in $g_2$.
We therefore obtain the explicit expression
\begin{equation}\label{g2}
g_2 (\rb \, ) \simeq  \rho_0(\rb) \rho_0 ({\bf 0}) 
+ \rho_0({\bf 0})\langle\delta\hat{\rho}(\rb)\rangle_3 
+ \rho_0(\rb)\langle\delta\hat{\rho}({\bf 0})\rangle_3 
+\langle \delta\hat{\rho}(\rb) \delta\hat{\rho}({\bf 0})\rangle_2
-\frac{\delta_{\rb,{\bf 0}}}{l^D} \rho_0({\bf 0}).
\end{equation}

This writing is however not the optimal one as the last term in $1/l^D$
gives the wrong impression that
$g_2(0)$ strongly depends on the discretisation length $l$ in
the continuous limit $l\rightarrow 0$.
In fact this strong dependence exactly compensates a term in $1/l^D$
in the density fluctuations $\< \dro^2 ({\bf 0}) \>$ coming from the fact that 
$\dro^2({\bf 0})$ is a product of field operators not in normal order.
To reveal this fact we express $\delta\hat{\rho}$ in terms of the operator
$\hat{\Lambda}$ of (\ref{devLam}) 
\begin{equation}
\delta\hat{\rho}(\rb)=\sqrt{\rho_0(\rb)}\left(\hat{\Lambda}(\rb)+\hat{\Lambda}^\dagger(\rb)
\right)+\hat{P}\partial_{N_0} \rho_0(\rb;N_0)
\label{n_rho}
\end{equation}
and we put the resulting expression in normal order with respect to the field $\hat{\Lambda}$
thanks to the bosonic commutation relation (\ref{commLam}): 
\begin{equation}
\label{no_rho}
\delta\hat{\rho}(\rb)\delta\hat{\rho}({\bf 0})=\,
:\!\delta\hat{\rho}(\rb)\delta\hat{\rho}({\bf 0})\!: +
\frac{\delta_{\rb,{\bf 0}}}{l^D} \rho_0({\bf 0})
-N_0 \phi_0^2(\rb)\phi_0^2({\bf 0})
\end{equation}
where $:\ :$ is the standard notation to represent normal order.
The spurious
term in $1/l^D$ is then exactly canceled:
\begin{eqnarray}
g_2(\rb)&=& N_0(N_0-1) \phi_0^2(\rb) \phi_0^2 ({\bf 0})
\nonumber \\
&+& \rho_0({\bf 0})\langle\delta\hat{\rho}(\rb)\rangle_3
+ \rho_0(\rb)\langle\delta\hat{\rho}({\bf 0})\rangle_3
+ \langle :\!\delta\hat{\rho}(\rb)\delta\hat{\rho}({\bf 0})\!:\rangle_2.
\label{g2:no}
\end{eqnarray}
This expression allows to prove the equivalence with the prediction for $g_2$
in the Bogoliubov theory. 
We do not present the calculations here, as they are a straightforward application
of appendix \ref{appen:CD}.
Finally we give a last alternative expression to $g_2$ equivalent to 
(\ref{g2:no}) at the present order:
\begin{equation}
\label{g2_alter}
g_2(\rb) = \left(1-1/N\right) \rho({\bf 0})\rho(\rb)
+ \langle :\!\delta\hat{\rho}(\rb)\delta\hat{\rho}({\bf 0})\!:\rangle_2
\end{equation}
where $N$ is the mean total number of particles
and $\rho$ is the mean total density:
\begin{equation}
\rho(\rb) = \rho_0(\rb) + \langle\delta\hat{\rho}(\rb)\rangle_3.
\label{densite_totale}
\end{equation}

\subsection{First order correlation function}\label{subsecg1}

The first order correlation function of the field is defined as:
\begin{equation}
g_1 (\rb) \equiv \< \psd (\rb) \ps ({\bf 0}) \>  
= \< \sqrt{\hat{\rho} (\rb \, )} e^{i (\tet ({\bf 0}) - \tet
(\rb \, ))} \sqrt{\hat{\rho} ({\bf 0}) } \>.
\end{equation}
As previously done we perform the calculation up to second order
in the small parameters $\epsilon_{1,2}$. We therefore expand 
$\sqrt{\hat{\rho}}$  up to second order in $\delta\hat{\rho}$
using (\ref{exp_rho}). Note that we do not expand the exponential in
$\tet ({\bf 0}) - \tet(\rb \, )$, contrarily to what
we did in the Hamiltonian: as $\rb$ and ${\bf 0}$ are not
neighboring points of the lattice anymore, the phase difference
of the field can be arbitrarily large.
The expansion in $\delta\hat{\rho}$ gives rise to six terms:
\begin{eqnarray}
g_1 (\rb) &=&\rho_0^{1/2}(\rb)\rho_0^{1/2}({\bf 0})
\left[\langle e^{i\Delta\theta}\rangle 
+\frac{1}{2} \langle
\delta\tilde{\rho}(\rb) e^{i\Delta\theta} + e^{i\Delta\theta} \delta\tilde{\rho}({\bf 0})
\rangle\right. \nonumber \\
&-& \left. \frac{1}{8} \langle \delta\tilde{\rho}^2(\rb)  e^{i\Delta\theta}
+ e^{i\Delta\theta} \delta\tilde{\rho}^2({\bf 0})
-2 \delta\tilde{\rho}(\rb) e^{i\Delta\theta} \delta\tilde{\rho}({\bf 0})
\rangle
\right]
\label{six_termes}
\end{eqnarray}
where we have introduced the following notations to simplify the writing:
\begin{eqnarray}
\Delta{\theta} &\equiv& \hat{\theta}({\bf 0}) - \hat{\theta}(\rb) \\
\delta \tilde{\rho}(\rb) &\equiv& \frac{\delta\hat{\rho}(\rb)}{\rho_0(\rb)} .
\end{eqnarray}
We calculate the expectation values in this expression in two steps,
first using the thermal equilibrium distribution for $H_2$, and
then including the corrections due to $H_3$.

The thermal expectation values corresponding to the quadratic Hamiltonian
$H_2$ are evaluated using Wick's theorem. One first expands the exponential
in powers of $\Delta\theta$, one calculates the expectation value
of each term, and then one performs an exact resummation of the resulting series. 
This leads to the simple identities
\begin{eqnarray}
\label{0op}
\langle e^{i\Delta\theta}\rangle_2 &=& e^{-\langle(\Delta\theta)^2\rangle_2/2} \\
\label{1op}
\langle \delta\tilde{\rho}(\rb) e^{i\Delta\theta}\rangle_2  &= &
e^{-\langle(\Delta\theta)^2\rangle_2/2} \langle \delta\tilde{\rho}(\rb)  
i\Delta\theta\rangle_2 
\\
\label{2op}
\langle\delta\tilde{\rho}^2(\rb) e^{i\Delta\theta}\rangle_2  &=&
e^{-\langle(\Delta\theta)^2\rangle_2/2}
\left[\langle \delta\tilde{\rho}^2(\rb)\rangle_2 
+\left(\langle\delta\tilde{\rho}(\rb) i\Delta\theta\rangle_2\right)^2
\right] \\
\label{1op1}
\langle \delta\tilde{\rho}(\rb) e^{i\Delta\theta} \delta\tilde{\rho}({\bf 0})\rangle_2 
&=& e^{-\langle(\Delta\theta)^2\rangle_2/2}
\left[\langle \delta\tilde{\rho}(\rb) \delta\tilde{\rho}({\bf 0})\rangle_2 + 
\langle \delta\tilde{\rho}(\rb) i\Delta\theta\rangle_2
\langle i\Delta\theta \delta\tilde{\rho}({\bf 0}) \rangle_2 \right] .
\end{eqnarray}
The expectation value of a product of the density fluctuation $\delta\hat{\rho}$
and of the phase
variation $i\Delta\theta$ is particularly simple. In a classical field theory
this expectation value would obviously vanish, as there is no crossed term
in $H_2$ between the density fluctuations and the phase. In the present 
quantum field theory this is not exactly the case as $\hat{\rho}$ and $\hat{\theta}$
do not commute. To show that, we use the fact that the Bogoliubov mode functions
$u_s(\rb)$, $v_s(\rb)$ can be chosen to be real,
so that $\delta\hat{\rho}$ and $i\Delta\theta$ are linear combinations
of $\hat{b}_s,\hat{b}_s^\dagger, \hat{P}$ with real coefficients. As
a consequence 
\begin{equation}
\langle \delta\tilde{\rho}(\rb) i\Delta\theta\rangle_2 = 
\langle \delta\tilde{\rho}(\rb) i\Delta\theta\rangle_2^*=  
-\langle i\Delta\theta \delta\tilde{\rho}(\rb)\rangle.
\end{equation}
This leads to
\begin{equation}
\langle \delta\tilde{\rho}(\rb) i\Delta\theta\rangle_2 = \frac{i}{2}
\langle  [\delta\tilde{\rho}(\rb), \Delta\theta]\rangle_2
= \frac{1-\delta_{\rb,{\bf 0}}}{2 \rho_0(\rb)l^D} .
\end{equation}
The same reasoning can be made for the other expectation value
\begin{equation}
\langle i\Delta\theta \delta\tilde{\rho}({\bf 0})\rangle_2 =
\frac{1-\delta_{\rb,{\bf 0}}}{2 
\rho_0({\bf 0}) l^D}.
\end{equation}
These expressions are second order in $\epsilon_{1,2}$. An important consequence
is that the product of such crossed phase-density expectation values 
in (\ref{2op},\ref{1op1}) are actually negligible at the present order
of the calculation. The resulting form for $g_1$, at the level of $H_2$,
is quite simple:
\begin{equation}
g_1(\rb)|_{H_2} =
\rho_0^{1/2}(\rb)\rho_0^{1/2}({\bf 0}) e^{-\langle(\Delta\theta)^2\rangle_2/2}
\left[1 -\frac{1}{8} \langle(\Delta\delta\tilde\rho)^2\rangle_2
+\frac{1}{4}\mbox{scoria}(\rb) \right]
\end{equation}
The notation $\Delta\delta\tilde{\rho}$ is similar to the one for the phase:
\begin{equation}
\Delta\delta\tilde{\rho}\equiv
\delta\tilde\rho({\bf 0})- \delta\tilde\rho(\rb)
\end{equation}
and the `scoria' comes from the crossed expectation value of $\Delta\theta$
and $\delta\tilde\rho$:
\begin{equation}
\mbox{scoria}(\rb)\equiv
\left(1-\delta_{\rb,{\bf 0}}\right)
\left(\frac{1}{\rho_0(\rb)l^D} + \frac{1}{\rho_0({\bf 0}) l^D} \right).
\end{equation}

At this point we face the same apparent problem as in the calculation of $g_2$:
the `scoria' scales as $1/l^D$ and gives the wrong impression that
our expression for $g_1$ will depend dramatically on $l$ in the
continuous limit $l\rightarrow 0$. As in the case of $g_2$, we solve this
problem by expressing $\delta\hat{\rho}$ and $\tet$ in terms
of the field $\hat{\Lambda}$ and putting the operators $\hat{\Lambda}$,
$\hat{\Lambda}^\dagger$ in normal order. We use (\ref{n_rho}) for the expression
of $\delta\hat{\rho}$. For the difference of two phase operators,
$\hat{Q}$ and the antihermitian operator $\hat{\alpha}$ cancel so that
\begin{equation}
\Delta\theta = \frac{1}{2i}\left(\Delta\tilde{\Lambda}-
\Delta\tilde{\Lambda}^\dagger\right)
\end{equation}
where we have introduced the notations
\begin{eqnarray}
\tilde\Lambda(\rb) &\equiv& \frac{\hat\Lambda(\rb)}{\rho_0^{1/2}(\rb)}\\
\Delta\tilde\Lambda &\equiv& 
\tilde\Lambda({\bf 0})- \tilde\Lambda(\rb).
\end{eqnarray}
After some calculations we arrive at
\begin{eqnarray}
\langle(\Delta\delta\tilde\rho)^2\rangle_2 &= &
\langle:\!(\Delta\delta\tilde\rho)^2\!:\rangle_2 + \mbox{scoria}(\rb) \\
\langle(\Delta\theta)^2\rangle_2 &=& 
\langle:\!(\Delta\theta)^2\!:\rangle_2 +\frac{1}{4}\mbox{scoria}(\rb).
\end{eqnarray}
As the `scoria' is second order in $\epsilon_{1,2}$, the exponential function
of it can be expanded to first order. We then find as expected that all
the $1/l^D$ terms exactly cancel:
\begin{equation}
g_1(\rb)|_{H_2} =
\rho_0^{1/2}(\rb)\rho_0^{1/2}({\bf 0}) 
e^{-\langle:(\Delta\theta)^2:\rangle_2/2}
\left[1 -\frac{1}{8} \langle:\!(\Delta\delta\tilde\rho)^2:\rangle_2 \right].
\label{g2H2}
\end{equation}

The last step is to include the first order correction 
to $g_1$ coming from the cubic Hamiltonian $H_3$.
One then has to calculate expectation values with 
the thermal equilibrium density 
operator $\exp[-\beta(H_2+H_3)]$ to first order in $H_3$.
This thermal density operator can be viewed as the evolution
operator during the imaginary time $-i\hbar\beta$ so that one
can use first order time dependent perturbation theory to get
\begin{equation}
\label{pert_form}
\langle \hat{O}\rangle_3 \simeq \langle \hat{O}\rangle_2 
-\int_0^{\beta} d\tau\, 
\langle  e^{\tau H_2} H_3 e^{-\tau H_2} \hat{O} \rangle_2
\end{equation}
where $\hat{O}$ is an arbitrary operator of the gas and where we have used the fact
that $H_3$ has a vanishing expectation value in the thermal equilibrium state for $H_2$.
One is back to the calculation of expectation values 
of some operators in the thermal state corresponding to $H_2$.
Wick's theorem can be applied. The resulting calculations
are very similar to the ones leading to (\ref{g2H2}),
but more involved and are detailed in the appendix \ref{appen:g2H3}.
The same phenomena takes place, that terms of a higher order
than the present calculation can be neglected. One then gets
\begin{eqnarray}
\label{id1}
\langle e^{i\Delta\theta}\rangle _3 &\simeq& 
e^{-\langle(\Delta\theta)^2\rangle_2/2}
\left[1 +\langle i\Delta\theta\rangle_3 \right] \\
\langle \delta\tilde\rho (\rb) e^{i\Delta\theta}\rangle_3 &\simeq&
e^{-\langle(\Delta\theta)^2\rangle_2/2}
\left[
\langle \delta\tilde\rho (\rb) i\Delta\theta \rangle_2
+\langle \delta\tilde\rho(\rb)\rangle_3 \right].
\label{id2}
\end{eqnarray}
The first terms in the right-hand side of (\ref{id1}) and (\ref{id2}) already appeared at the
level of $H_2$, the second terms are corrections due to $H_3$ that we now take into account.
There is no need to include $H_3$ corrections to the other terms
of (\ref{six_termes}) since they are quadratic in $\delta\hat{\rho}$
and are therefore already of second order.
The expectation values of the phase $\hat{\theta}$ 
and of the density fluctuations $\delta\hat{\rho}$ have been 
calculated in subsection \ref{cubic_corr}. It was found that
the expectation value of the phase operator is space independent
so that $\langle \Delta\theta\rangle_3$ vanishes.
The expectation value of the density fluctuations
including the effect of $H_3$ 
was given in (\ref{resultat}) and is in general different from zero.
Remarkably the whole effect on the correlation
function $g_1$ of the first order correction
in $H_3$ is to replace $\rho_0(\rb)$ by
the total mean density $\rho(\rb)$ defined in (\ref{densite_totale})!

We write our final expression for the first order correlation function
of the field, calculated consistently up to $\epsilon_{1,2}^2$:
\begin{equation}
g_1(\rb)=\sqrt{\rho(\rb)\rho({\bf 0})}
\exp\left[-\frac{1}{2}\langle:(\Delta\theta)^2:\rangle_2
-\frac{1}{8} \langle:\!(\Delta\delta\tilde\rho)^2:\rangle_2 \right].
\label{g1_jolie}
\end{equation}
Note that we have inserted the contribution of the density fluctuations
inside the exponential factor, which is allowed at the order of the present
calculation since this contribution is of order $\epsilon_{1,2}^2$.

What happens in the regime where a true condensate is present~?
Both phase and density fluctuations are small, so that the
exponential function in (\ref{g1_jolie}) can be expanded
to first order. We then express $\Delta\theta$ and $\Delta\tilde{\rho}$
in terms of the operator $\hat{\Lambda}$ and of the operator $\hat{P}$.
Since the Bogoliubov theory is usually considered in the canonical
ensemble we remove the terms corresponding to the grand canonical
fluctuations of the particle number.
We then recover exactly the Bogoliubov prediction:
\begin{equation}\label{smallphase}
g_1^{\rm Bog}(\rb) = \Psi_c (\rb) \Psi_c ({\bf 0}) + 
\< \hat{\Lambda}^\dagger (\rb \,) \hat{\Lambda} ({\bf 0}) \>
\end{equation}
where $\Psi_c(\rb) = \sqrt{N_0}\phi_0(\rb)+\chi(\rb)/\sqrt{N_0}$
is the condensate field.
Amazingly the general formula (\ref{g1_jolie}) 
for quasi-condensates can be related to the Bogoliubov formula
in the following very simple way:
\begin{equation}
g_1(\rb)=\sqrt{\rho(\rb)\rho({\bf 0})}
\exp\left[\frac{g_1^{\rm Bog}(\rb)}{\sqrt{\rho(\rb)\rho({\bf 0})}}-1
\right].
\label{g1_lwb}
\end{equation}

\section{Explicit results for the spatially homogeneous case}\label{homogene}
In this section we apply our approach to a spatially homogeneous Bose gas.
The quasi-condensate density is then uniform:
\begin{equation}
\rho_0(\rb) = \frac{N_0}{L^D}=\frac{\mu}{g_0}.
\end{equation}
The Bogoliubov equations
(\ref{Bogo}) can then be exactly solved for any dimension of space and lead to 
$u_k (r) = \bar{u_k} \, e^{i k r} / L^{D/2}$ 
and $v_k (r) = \bar{v_k} \, e^{i k r} / L^{D/2}$ with:
\begin{equation}\label{1Dfluc}
\bar{u_k} - \bar{v_k}  =   \Big \lbrack \frac{ \frac{\hbar^2 k^2}{2 m} + 2 \mu }
{ \frac{\hbar^2 k^2}{2 m} } \Big \rbrack^{1/4} \qquad 
\textrm{and} \qquad \bar{u_k} + \bar{v_k}  =   
\Big \lbrack \frac{ \frac{\hbar^2 k^2}{2 m} }{ \frac{\hbar^2 k^2}{2 m} + 2 \mu }
 \Big \rbrack^{1/4}.
\end{equation}
The corresponding eigenenergies are given by: 
\begin{equation}
\epsilon_k=  
\Big \lbrack \frac{\hbar^2 k^2}{2 m} (\frac{\hbar^2 k^2}{2 m} + 2 \mu) \Big \rbrack^{1/2}.
\end{equation}

\subsection{Equation of state}
 From the general expression (\ref{muC}) for the chemical potential of the gas
 we arrive in the thermodynamical limit at
\begin{equation}\label{muC2}
\mu = \rho g_0 + g_0 \int_{\mathcal{D}} \frac{d\kb}{(2 \pi)^D}
[(\bar{u}_k+\bar{v}_k)^2 n_k+\bar{v}_k (\bar{u}_k +\bar{v}_k)]
\end{equation}
where $n_k=1/(\exp(\beta\epsilon_k)-1)$ is the mean occupation number
of the Bogoliubov mode $\kb$. 
$\mathcal{D} = [-\pi/l,\pi/l]^D$ is the square domain of integration in 
the ${\bf k}$ space. 
The integral over the wavevector ${\bf k}$ does not contain  any infrared 
divergence for any dimension of space. However the long wavevector behavior 
given by 
\begin{equation}\label{longwave}
\bar{v}_k (
\bar{u}_k +\bar{v}_k ) \simeq - \frac{m \mu}{\hbar^2 k^2}
\end{equation}
gives rise to an integral
 convergent in $1D$ and divergent in $2D$ and $3D$ in the $l \to 0$ limit.
This gives the impression that the result depends strongly on $l$.
The solution of this paradox comes from the link between the bare coupling
constant $g_0$ of the model potential in the discretized space and the low
energy two-body scattering properties of the exact potential in the 
continuous space. 
 This gives to $g_0$ in two and three dimensions a dependence in $l$ 
so that our expression for $\mu$ does not depend on $l$ anymore in the 
$l \to 0$  limit.
In one dimension, the bare coupling $g_0$ is simply equal to the actual 
coupling strength $g$ for $l \to 0$ and there is no divergence. A $T=0$, 
the equation (\ref{muC2}) leads to:
\begin{equation}
\mu = g \rho \left( 1 - \frac{1}{\pi \rho \, \xi} \right)
\end{equation}
where $\xi$ is the healing length defined in (\ref{def_xi}).
This agrees with the result of Lieb-Liniger in the weak interaction
limit \cite{lieb}.
In three dimensions, we 
refer to the appendix of \cite{cartago} where the calculation has been done. 
One finds:
\begin{equation}\label{correc1}
g_0 = \frac{g}{1 - \displaystyle  g\int_{\mathcal{D}} \frac{d\kb}{(2 \pi)^3}
\frac{1}{\hbar^2 k^2/m}}.
\end{equation} 
 $g$ is the usual $3D$ coupling 
strength given by :
\begin{equation}
g = \frac{4 \pi \hbar^2 a}{m}
\end{equation}
where $a$ is the exact potential scattering length. A more explicit form
of (\ref{correc1}) is
\begin{equation}
g_0 = \frac{g}{1-K a/l}
\end{equation}
where $K=2.442\ldots$.
It has to be noted that 
the difference between $g_0$ and $g$ is still small in the validity domain of our
approach since it is a second 
order correction in $\epsilon_{1,2}$: taking $l\sim \xi$ one finds
$a/l\sim 1/\rho l^3$. Replacing the first
factor $g_0$  in (\ref{muC2}) with 
the formula (\ref{correc1}) expanded
up to second order in $\epsilon_{1,2}$ gives:
\begin{equation}
\mu = \rho g +g_0 \int_{\mathcal{D}} \frac{d\kb}{(2 \pi)^3} \left(
(\bar{u}_k+\bar{v}_k)^2 n_k+\bar{v}_k (
\bar{u}_k +\bar{v}_k ) + \frac{m \mu}{\hbar^2 k^2} \right)
\end{equation}
One can then safely take the $l \to 0$ limit. At $T=0$, the integration gives:
\begin{equation}
\mu= g \rho \left( 1+ \frac{32 \sqrt{\pi}}{3} \sqrt{\rho a^3} \right)
\end{equation}
which is the same result as Lee and Yang's \cite{Leeyang}.
In two dimensions, the low energy two-body scattering of a general short
range potential is  described by a 
single length $a$  also named the scattering length. In a continuous space, 
the $T$ matrix can be calculated in the low energy limit:
\begin{equation}
\label{lel}
\langle {\bf k} | T(E+i \eta) | {\bf k'} \rangle \simeq - 
\frac{2 \pi \hbar^2}{m \left[ \ln \left( \frac{a k_0}{2} \right) + C - i 
\pi /2 \right]}
\end{equation}
where $C = .57721\ldots$ is the Euler constant, $a$ is the 
scattering length, $E = \hbar^2 k_0^2 / m$ and 
$\eta \to 0^{+}$. 
We can also calculate the $T$ matrix for the discrete delta potential defined
by Eq. (\ref{potential}) which can also be expressed as:
\begin{equation}
\label{potential2}
V = \frac{g_0}{l^2}  | { \bf r} = 0 \rangle  \langle { \bf r} = 0 |.
\end{equation}
The general scattering theory gives  the relations between the $T$ matrix, 
the propagator $G$ and the free  propagator $G_0$:
\begin{eqnarray}
T &=& V + V G V \\
G &=& G_0 + G_0 V G 
\end{eqnarray}
Using these relations and Eq. (\ref{potential2}) for the potential, we find:
\begin{equation}
\langle {\bf k} | T_{\rm grid}(E+i \eta) | {\bf k'} \rangle = 
 \frac{g_0}{1-g_0 \langle { \bf r} = 0 | G_0(E+i \eta) | { \bf r} = 0 
\rangle}.
\end{equation}
The only term we need to calculate is the free 
propagator taken at the origin which is conveniently performed with a Fourier
transform:
\begin{equation}
\langle { \bf r} = 0 | 
G_0(E+i \eta) | { \bf r} = 0 \rangle = \int_{\mathcal{D}}
 \frac{d\kb}{(2 \pi)^2} \frac{1}{E+i\eta - 
\hbar^2 k^2/m}.
\end{equation}
We split the square $\mathcal{D}$ into a disk of radius $\pi / l$ and the
complementary domain. Integration over the complementary domain gives simply
a constant term in the low energy limit $E \ll \hbar^2 / m l^2$:
\begin{equation}\label{intedomain}
J \equiv \frac{2 \pi \hbar^2}{m} \int_{\mathcal{D} - {\rm disk}} 
\frac{d\kb}{(2 \pi)^2} 
\frac{1}{E+i\eta 
- \hbar^2 k^2/m} \simeq - \frac{1}{2 \pi} \int_{\mathcal{D} 
- {\rm disk}} 
\frac{d\kb}{k^2} =   \frac{2 G}{\pi} - 
\ln(2) 
\end{equation}
where $G = .91596$ is the Catalan constant.  The disk integration is 
straightforward and leads to the following expression for the $T$ matrix:
\begin{equation}
\langle {\bf k} | T_{\rm grid}(E+i \eta) | {\bf k'} \rangle =
\frac{1}{\displaystyle
\frac{1}{g_0} - \frac{m}{2 \pi \hbar^2} \ln \left(\frac{l k_0}{\pi}
\right) + \frac{i m}{4 \hbar^2} - \frac{m}{2 \pi \hbar^2} J}.
\end{equation}
We now take $T_{\rm grid} = T$, where $T$ is approximated by 
(\ref{lel}), in order to reproduce the 
low-energy scattering properties of the exact potential. This leads to:
\begin{equation}\label{g0final}
\frac{1}{g_0} = \frac{m}{2 \pi \hbar^2} \left( \ln \left( \frac{ l}{\pi a} 
\right) - C + \frac{2 G}{\pi} \right).
\end{equation}
Note that the condition (\ref{inter}) has to be satisfied in our approach. 
In two dimensions, this gives $\hbar^2/m g_0 \gg 1$ or using 
Eq. (\ref{g0final}):
\begin{equation}
\frac{1}{2\pi} \ln \left( \frac{l}{a} \right) \gg 1.
\end{equation}
We now show that the logarithmic
dependence in $l$ appearing in  $g_0$, Eq. (\ref{g0final}),
exactly cancels 
the one appearing in the equation of state. Eq. (\ref{muC2}) can be 
rewritten as:
\begin{equation}\label{rho}
\rho = \frac{\mu}{g_0}   -   \int_{\mathcal{D}} 
\frac{d\kb}{(2 \pi)^2}  \left[ ( \bar{u_k} + \bar{v_k} )^2  
n_k + \bar{v_k} (  \bar{u_k} + \bar{v_k} ) \right].
\end{equation}
In the thermal part, one can immediately take the $l \to 0$ limit.  In order 
to calculate the integral corresponding to the $T=0$ case, we use the 
same technique as for the calculation of $g_0$: the integration is made on a 
disk domain of radius $\pi/l$ and we keep as a correction
the  integration over the complementary  domain. The complementary domain 
integration is made by using the high wavevector behavior of $\bar{v_k} 
(  \bar{u_k} + \bar{v_k} )$, Eq. (\ref{longwave}).
 This leads to:
\begin{equation}\label{deltaNfinal}
 - \int_{ \cal D} \frac{d\kb}{(2 \pi)^2} 
\bar{v_k} ( \bar{u_k} + \bar{v_k}) = \frac{m \mu}{4 \pi \hbar^2}\left[\ln \left( 
\frac{\pi^2 \hbar^2}{m l^2 \mu} \right)-1-2 J\right].
\end{equation}
Using Eqs.(\ref{g0final}) and (\ref{deltaNfinal}) in Eq. (\ref{rho}), we 
arrive at an implicit equation of state:
\begin{equation}\label{rho2}
\rho = \frac{m \mu}{4 \pi \hbar^2} \ln \left( \frac{4 \hbar^2}{a^2 m \mu 
e^{2 C +1}} \right) -  \int
\frac{d\kb}{(2 \pi)^2}  ( \bar{u_k} + \bar{v_k} )^2  
n_k .
\end{equation}
Remarkably this is identical to the result (20.45) obtained by the
functional integral method in \cite{popov}.
A $T=0$, one can show from the condition $\rho \xi^2
\gg 1$, see Eq.(\ref{inter}), that the validity condition of our approach is
$\ln ( 1 / \rho a^2 ) \gg 4\pi$.
If one approximately inverts Eq.(\ref{rho2}) neglecting constant terms and 
$\ln( \ln( 1/ \rho a^2))$ with respect to $\ln ( 1 / \rho a^2 )$, one recovers
Schick's formula \cite{schick}.

\subsection{Are density and gradient-of-phase fluctuations small~?}
As mentioned in section \ref{Hquadra}, our approach relies in particular on
two assumptions: the assumption that the relative density fluctuation 
$\epsilon_1$ is small, and the assumption that the phase variation 
$\epsilon_2$ between two neighboring points of the grid 
is small.

Let us consider first the relative density fluctuations. Thanks to (\ref{no_rho})
their mean square value can be separated in two parts:
\begin{equation}
\label{fluc_dens}
\epsilon_1^2=\frac{\langle\delta\hat{\rho}^2({\bf 0})\rangle_2}{\rho_0^2}=\,
\frac{1}{\rho_0l^D} +
\frac{\langle:\!\delta\hat{\rho}^2({\bf 0})\!:\rangle_2}{\rho_0^2} 
\end{equation}
where we have neglected $1/L^D$ with respect to $1/l^D$ in the thermodynamical limit.
The second term in (\ref{fluc_dens}), 
involving the normal order, is expressed in terms of the 
$\bar{u}_k,\bar{v}_k$ in the thermodynamical limit as
\begin{equation}
\label{second_terme}
\frac{\langle:\!\delta\hat{\rho}^2({\bf 0})\!:\rangle_2}{\rho_0^2} =\frac{2}{\rho_0}
\int_{\cal D} \frac{d\kb}{(2\pi)^D}
\left[(\bar{u}_k+\bar{v}_k)^2 n_k + \bar{v}_k(\bar{u}_k+\bar{v}_k)\right] 
\end{equation}
where the integration domain is ${\cal D}=[-\pi/l,\pi/l]^D$.
At zero temperature one introduces the change of variable ${\bf q} = \kb \xi$
in the integral: one finds that (\ref{second_terme}) is of the order of
$1/\rho_0\xi$ in 1D, of the order of $\ln(\xi/l)/\rho_0\xi^2$ in 2D
and of the order of $1/\rho_0 \xi^2 l$ in 3D. Since $l<\xi$
the second term in (\ref{fluc_dens}) is dominated by the first term and one
has indeed
\begin{equation}
\label{indeed}
\epsilon_1^2 \simeq \frac{1}{\rho_0 l^D}.
\end{equation}
At finite temperature we have to calculate the thermal contribution to
(\ref{second_terme}) involving the occupation number $n_k$. 

At a temperature
$k_B T < \mu$ we use the low momentum expansion of the $\bar{u}_k+\bar{v}_k$
and $\epsilon_k$ and we find that the thermal contribution is
$(k_B T/\mu)^{D+1}(l/\xi)^D$ times smaller than $1/\rho_0 l^D$.

At a temperature $k_B T >\mu$, that is $\lambda < \xi$,
the treatment depends on the dimension of space. 
In 1D the main contribution to the integral comes
from the domain $\epsilon_k \sim \mu$, over which one can approximate
the Bose formula by its low energy limit $k_B T/\epsilon_k$.
This leads to a normal ordered fluctuation (\ref{second_terme}) of the
order of $k_B T/(\mu \rho_0 \xi)$. This is larger than $1/\rho_0 \lambda$
so that the condition $l<\lambda$ then no longer implies
that the first term $1/\rho_0 l$ in (\ref{fluc_dens}) is the dominant 
one. For convenience one can however adjust $l$ to a value such that 
\begin{equation}
\frac{1}{\rho_0 l} \sim \frac{k_B T}{\mu} \, \frac{1}{\rho_0\xi}.
\end{equation}
The condition for weak density fluctuations then becomes
\begin{equation}
\epsilon_1^2 \sim  \frac{k_B T}{\mu} \, \frac{1}{\rho_0\xi} \ll 1.
\end{equation}
Using $\rho_0\simeq\rho$ and $\mu \simeq g \rho$ we recover a condition
already obtained in \cite{yvan1} with a pure classical field
approach. Note that this condition can be rewritten as $\xi \ll l_c$
where the coherence length of the field will be defined in (\ref{asymp}).
In 2D both the low energy domain $\epsilon_k < k_B T$ and the high
energy domain $\epsilon_k > k_B T$ have an important contribution.
In the low energy domain we approximate the Bose law by its low energy
limit. In the high energy domain we keep the full Bose law but $\epsilon_k$
being then larger than $\mu$, we approximate $\bar{u}_k+\bar{v}_k$ by 
unity and $\epsilon_k$ by $\hbar^2 k^2/2m$. This leads to a normal 
ordered fluctuation (\ref{second_terme}) of the 
order of $\ln(k_B T/\mu) k_B T/(\mu \rho_0 \xi^2)$,
a quantity which is larger than $1/\rho_0 \lambda^2$. 
As in 1D we therefore adjust $l$ so that
\begin{equation}
\epsilon_1^2 \sim \frac{1}{\rho_0 l^2} \sim \frac{k_B T}{\mu} 
\ln\left(\frac{k_B T}{\mu}\right)
\, \frac{1}{\rho_0\xi^2}.
\end{equation}
In 3D the high energy domain $\epsilon_k > k_B T$ gives the dominant 
contribution so that the normal ordered expectation value (\ref{second_terme})
scales as $1/\rho_0 \lambda^3$. This is dominated by the first term
in (\ref{fluc_dens}) so that the estimate (\ref{indeed}) applies
as soon as $l<\lambda,\xi$.

Let us consider now the condition that the mean squared phase change over a grid cell:
\begin{equation}
\label{mgrad}
\epsilon_2^2 = \langle (l\nabla\hat{\theta})^2\rangle_2= \frac{l^2}{2\rho_0}
\int_{\cal D} \frac{d\kb}{(2\pi)^D}
k^2 (\bar{u}_k-\bar{v}_k)^2 (n_k+1/2)
\end{equation}
is much smaller than unity.
The presence of the factor $k^2$ inside the integral, coming
from the action of $\nabla$, has the consequence that the
contribution to the integral is dominated by the high energy
domain.
At zero temperature one can replace $\bar{u}_k-\bar{v}_k$ by unity
since the integral is dominated by wavevectors $k\sim 1/l > 1/\xi$. 
This leads to
\begin{equation}
\epsilon_2^2 \sim \frac{1}{\rho_0 l^D}
\label{eps2bis}
\end{equation}
whatever the dimension $D$.

At a temperature $k_B T < \mu$ we estimate the thermal contribution by
replacing $\bar{u}_k-\bar{v}_k$ and $\epsilon_k$ by their low momentum
approximations: the thermal contribution is then $(l/\lambda)^{2+D}
(\xi/\lambda)^D$ times smaller than the zero temperature result (\ref{eps2bis})
and is therefore negligible since $l<\xi<\lambda$.

At a temperature $k_B T > \mu$ we use the high energy approximation
replacing $\bar{u}_k-\bar{v}_k$ by unity and $\epsilon_k$ by $\hbar^2k^2/2m$.
Note that this works even in 1D thanks to the presence of the $k^2$
factor in the integral (\ref{mgrad}). This leads to a thermal
contribution which is $(l/\lambda)^{2+D}$ times smaller than
the zero temperature contribution (\ref{eps2bis}), and which is 
negligible since $l<\lambda$.

We conclude that the small parameter $\epsilon_2$ of the theory,
ensuring that there is a weak phase variation over a grid cell, is
always given by (\ref{eps2bis}) provided that the conditions 
(\ref{restriction}), (\ref{temp}) and (\ref{inter}) are satisfied.

One may wonder if the condition that the corrections of the mean density
due to the interaction $H_3$ between the Bogoliubov modes
lead to an extra validity condition of our treatment.
For the considered case of a spatially homogeneous gas
it turns out that the answer to this question is no. One has indeed
the remarkable identity in the thermodynamical limit:
\begin{equation}
\frac{1}{\rho_0} \langle\delta\hat{\rho}(\rb)\rangle_3=-\frac{1}{2\rho_0^2}
\langle : \delta\hat{\rho}(\rb)^2:\rangle_2.
\end{equation}
If the relative density fluctuations are weak, 
the relative correction to the density will be also weak.

To end this subsection we discuss briefly the second order
correlation function of the field $g_2(\rb)$.
Restricting the general formula (\ref{g2_alter}) to the spatially 
homogeneous case in the thermodynamical limit we obtain
\begin{equation}
g_2(\rb) = \rho^2 +
2\rho \int_{\cal D} \frac{d\kb}{(2\pi)^D}    
\left[(\bar{u}_k+\bar{v}_k)^2 n_k + \bar{v}_k(\bar{u}_k+\bar{v}_k)\right]
\cos(\kb\cdot\rb).
\end{equation}
Limiting cases of this general formula can be compared to existing
results in the literature.
At zero temperature for a 1D Bose gas one gets for $r=0$:
\begin{equation}
\label{g20}
g_2(0) = \rho^2\left(1-\frac{2}{\pi\rho\xi}\right).
\end{equation}
This formula can be checked from \cite{lieb}: the mean interaction energy per
particle $v$ is equal to $g_2(0)$ multiplied by $g/2\rho$, and 
$v$ can be calculated in the weakly interacting
regime by combining (3.29) of \cite{lieb} (relating $v$ to the derivative
of the ground state energy with respect to $g$) and (4.2) of \cite{lieb}
(giving the ground state energy in the Bogoliubov approximation).
This exactly leads to (\ref{g20}).   This prediction for $g_2(0)$
also appears in \cite{shlyap_gangardt}.

\subsection{First order correlation function}
Thanks to the general formula (\ref{g1_lwb}) the first order correlation
function of the field for the quasi-condensate is immediately related
to the one of the Bogoliubov theory, here in the thermodynamical limit:
\begin{equation}
\ln[g_1(\rb)/\rho]=\frac{g_1^{\rm Bog} (\rb)}{\rho}
-1 =  
-\frac{1}{\rho}\int \frac{d^Dk}{(2\pi)^D}
\left[\left(\bar{u}_k^2+\bar{v}_k^2\right)n_k +\bar{v}_k^2 \right]
(1-\cos\bf{k}\cdot\rb).
\end{equation}
We have also taken here the continuous limit $l\to 0$, which does not lead
to any divergence.

We concentrate our analysis to the 1D case and we make the link with
existing results in the literature. These existing results deal with
the asymptotic behavior of $g_1$ for large $r$, where $r$ is the absolute
value of the spatial coordinate. At zero temperature,  we find
for $r\gg \xi$:
\begin{equation}
g_1 (r) \simeq \rho \: \left( \frac{r_1}{r} \right)^{1/ 2 \pi \rho \, \xi}
\end{equation}
with $r_1=e^{2-C} \, \xi /4 \simeq 1.037 \: \xi $
where $C=0.57721\ldots$ is Euler's constant \cite{note_ci}.
This reproduces a result obtained in a non explicit way by \cite{popovR}.
At a finite temperature, $g_1/\rho$ is the exponential of an integral
of the form
\begin{equation}
\int_0^{+\infty} 
\frac{A(k)}{k^2}[1-\cos(kr)]
\end{equation}
where the function $A(k)$ is a regular and even function of $k$, therefore
behaving quadratically with $k$ around $k=0$ \cite{note_A}. Writing $A(k)$ as $[A(k)
-A(0)]+A(0)$ and splitting accordingly the integral one obtains
for $r$ much larger than both $\xi$ and $\lambda$:
\begin{equation}
\label{asymp}
\ln[g_1(r)/\rho] = \frac{r}{l_c} + K + o(1/r^n)
\end{equation}
where the coherence length $l_c=\rho \lambda^2/\pi$ coincides with
the one of \cite{schwartz2} and the constant $K$ is given by
\begin{equation}
K=\int_0^{+\infty} \frac{A(k)-A(0)}{k^2}.
\end{equation}
Since $A(k)$ is even one can show by repeated integration by parts that
the remainder in (\ref{asymp}) tends to 0 faster than any power
law, contrarily to what is stated in \cite{schwartz2}.

\begin{figure}[h]
\includegraphics[width=0.8\textwidth,height=0.35
\textheight]{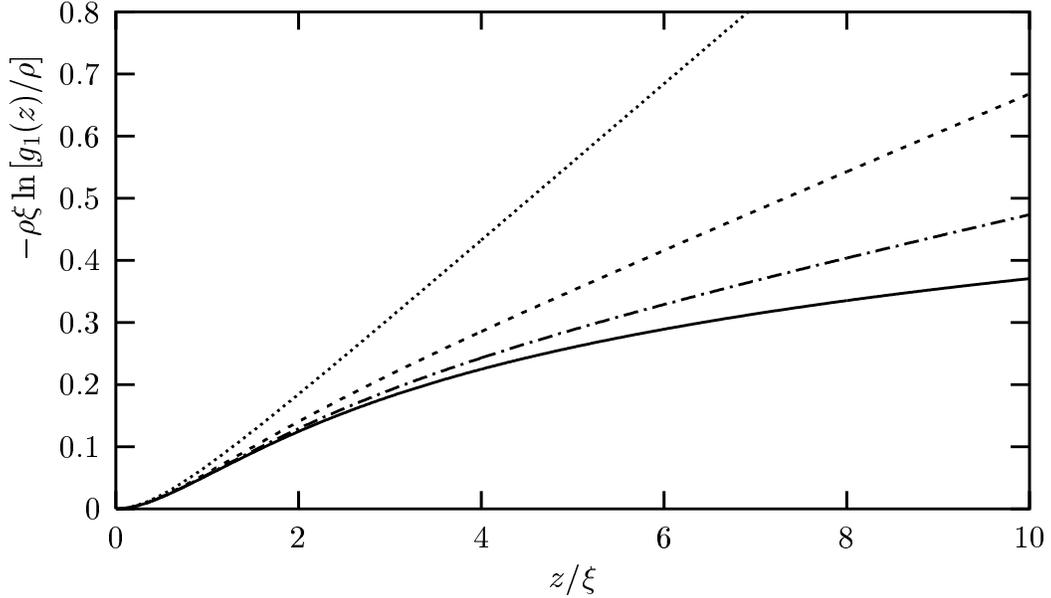}
\caption{First order correlation function of the field $g_1(z)$ for 
a repulsive 1D Bose gas in the thermodynamical limit.
The different curves correspond
to various ratios of the temperature to the chemical
potential: $k_B T/\mu=0$ (solid lines), $1/15$ (dot-dashed lines), 
$1/8$ (dashed lines),
$1/4$ (dotted lines). We plot the logarithm of $g_1(z)$ multiplied 
by the parameter $\rho \xi$, where $\rho$ is the 1D spatial density and $\xi=
\hbar/\sqrt{m \mu}$ is the healing length, so that we obtain a quantity 
depending
only on $z/\xi$ and $k_BT/\mu$ in the weakly interacting limit.
\label{fig1}}
\end{figure}

\begin{figure}[h!]
\includegraphics[width=0.7\textwidth,height=0.3
\textheight]{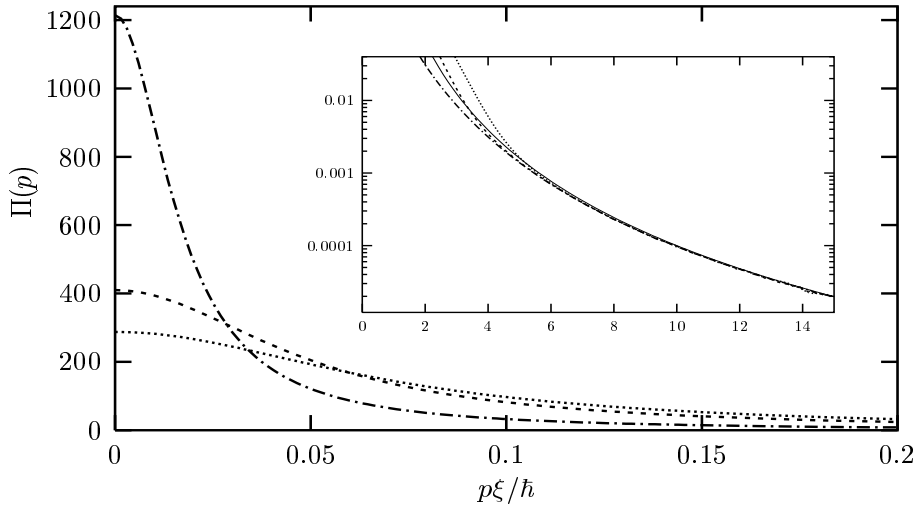}
\caption{ Momentum distribution of a repulsive 1D Bose gas in the 
thermodynamical limit. 
$\Pi(p)$ is normalized as $\int dp\Pi(p)=2\pi\hbar\rho$ where $\rho$ is
the 1D spatial density so that
$\Pi(p)$ is dimensionless.
We plot $\Pi(p)$ as a function of $p \xi/\hbar$
for various ratios of the 
temperature to the chemical
potential: $k_B T/\mu=1/3$ (dot-dashed lines), $1$ (dashed lines), 
$10/7$ (dotted lines). We have taken $\rho \xi =10 \gg 1$ where
$\xi= \hbar/\sqrt{m \mu}$ is the healing length.
 The solid line is the large 
$p$ limit: $(\hbar/p \xi)^4$.
\label{fig2}}
\end{figure}

Of course our formula gives access to $g_1$ for any value of
the distance.  This is illustrated in figure~\ref{fig1}
where we have plotted the logarithm of $g_1$ as function of $r/\xi$
for various temperatures.
As a consequence we can for example calculate the momentum distribution
of the atoms:
\begin{equation}
\Pi(p) = 2\int_0^{+\infty} dr\, g_1(r)\cos(pr/\hbar)
\end{equation}
normalized here as $\int dp\Pi(p)=2\pi\hbar\rho$ so that
$\Pi(p)$ is dimensionless. This is illustrated
in figure~\ref{fig2} where we have plotted the momentum distribution
for various temperatures  and
for $\rho\xi=10$.
Using integration by part we can show that the behavior 
of $\Pi$ for large $p$
is related to the fact that the third order derivative
of $g_1$ in $r=0^+$ does not vanish:
\begin{equation}
\Pi(p) \sim \frac{2\hbar^4g_1^{(3)}(0^+)}{p^4} 
\ \ \mbox{with}\ \ g_1^{(3)}(0^+)= \mu^2 m^2 / (2 \hbar^4).
\end{equation}
This prediction, valid at zero or finite temperature, 
agrees with the weak interaction limit of a recently obtained exact result
based on the Bethe ansatz \cite{olshanii}. At zero temperature we find
that the momentum distribution diverges in $p=0$ as
\begin{equation}
\Pi(p) \sim
\frac{\hbar\rho\pi\nu}{p}(r_1 p/\hbar)^{\nu}
\end{equation}
where $\nu=1/(2\pi\rho\xi)\ll 1$.

\section{Conclusion}

We have studied the thermal equilibrium
of weakly interacting degenerate Bose gases in the 
regime of weak density fluctuations, the so-called quasi-condensate
regime. The method can be considered as a Bogoliubov method
in the density-phase representation of the field operator.

In a first step one discretizes the real space in cells of size $l$:
$l$ is small enough that the macroscopic properties of the gas are not
affected by the discretisation, and $l$ is large enough that each
cell contains on the average a large number of particles.
The macroscopic occupation of each cell allows to give a precise
definition of the phase operator, as pioneered by Girardeau
\cite{girardeau}.

In a second step one performs a systematic expansion of the full Hamiltonian
in terms of two small parameters, the relative density fluctuations inside
a cell and the 
phase change over a grid cell.  This procedure
leads to an exact expansion of the observables of the gas in the regime of
weak interactions and low density fluctuations, in 1D, 2D and
3D.  In particular it is
free of any ultraviolet or infrared divergences and exactly
matches the usual Bogoliubov predictions when the gas contains a true
Bose-Einstein condensate.

As a first application of the general formalism, we have given
in this paper formulas for the equation of state of the gas, 
the ground state energy, the first order and second order 
correlation functions of the field. We have applied these formulas to
the spatially homogeneous case in 1D, 2D and 3D, recovering
in this way known results, but obtaining also to our knowledge new results,
like the full position dependence
of the first order correlation function of the field.

\acknowledgments
We acknowledge useful discussions with Iacopo Carusotto,
Gora Shlyapnikov, Dimitri Gangardt, Gordon Baym.
Laboratoire Kastler Brossel is a research unit of Ecole normale sup\'erieure and
Universit\'e Paris 6, associated to CNRS.

\appendix
\section{Expansion of the Hamiltonian}
\label{appen:expansion}
As explained in \ref{Hquadra} we expand the Hamiltonian (\ref{grandH}) up to third order
in powers
of the small parameters $\epsilon_1$ and $\epsilon_2$ defined in (\ref{eps1}), (\ref{eps2}).
This will produce terms $H^{(n_1,n_2)}$ of order $\epsilon_1^{n_1}\epsilon_2^{n_2}$
with $n_1+n_2\leq 3$.
The expansion of the potential energy part $H_{\rm pot}$ defined in (\ref{pot}) is very simple
as it only involves the operator giving the density. The only point is to realize that
the term $1/l^D$ is $\epsilon_1^2$ times smaller than the zeroth order density 
$\rho_0$. This leads to
\begin{eqnarray}
H_{\rm pot}^{(0,0)} &=& \sum_{\rb} l^D \rho_0 \left[U(\rb)-\mu+\frac{g_0}{2}\rho_0\right]  \\
H_{\rm pot}^{(1,0)} &=& \sum_{\rb} l^D \delta\hat{\rho} \left[U(\rb)-\mu+g_0
\rho_0\right]\\
H_{\rm pot}^{(2,0)} &=& \sum_{\rb} l^D \frac{g_0}{2}\left[\delta\hat{\rho}^2-\frac{\rho_0}{l^D}\right]. \\
H_{\rm pot}^{(3,0)} &=& - \sum_{\rb} \frac{g_0}{2} \delta\hat{\rho}.
\end{eqnarray}

The expansion of the kinetic energy part (\ref{hkin}) is more complicated as it involves also
the phase operator $\hat{\theta}$, which furthermore does not commute with $\delta\hat{\rho}$.
An expression slightly more convenient than (\ref{hkin}) can be given for the
kinetic energy. Thanks to the periodic boundary conditions one can freely
shift the summation variable $\bf$ in the term of (\ref{hkin}) involving
$\hat{\rho}_{-j}$, so that 
\begin{equation}
H_{\rm kin} = -\frac{\hbar^2}{2ml^2}\sum_{\rb,j} l^D\,  \left\{
\left[\sqrt{\hat{\rho}} e^{i(\hat{\theta}_{+j}-\hat{\theta})}\sqrt{\hat{\rho}_{+j}}
+\mbox{h.c.}\right]-2\sqrt{\hat{\rho}}\right\}.
\end{equation}
The calculation to zeroth order in $\epsilon_2$ can be done first easily: using
the expansion (\ref{exp_theta}) to zeroth order, we get from (\ref{hkin}) to
all orders in $\epsilon_1$:
\begin{equation}
H_{\rm kin}^{(\leq +\infty, 0)}  = -\frac{\hbar^2}{2m}
\sum_{\rb} l^D \, \sqrt{\hat{\rho}}\Delta\sqrt{\hat{\rho}}.
\end{equation}
It involves a function of $\hat{\rho}$ only  that it is easily expanded in powers of $\epsilon_1$
using (\ref{exp_rho}).
A simplification occurs after summation over the lattice, as the matrix $\Delta$
is symmetric for the considered periodic boundary conditions:
\begin{equation}
\sum_{\rb} u\Delta v  = \sum_{\rb}(\Delta u)\, v
\end{equation}
where $u$ and $v$ are arbitrary functions on the lattice. This leads to
\begin{eqnarray}
H_{\rm kin}^{(0,0)} &=&-\frac{\hbar^2}{2m} l^D \sum_{\rb} \sqrt{\rho_0}\Delta\sqrt{\rho_0} \\
H_{\rm kin}^{(1,0)} &=& -\frac{\hbar^2}{2m} l^D \sum_{\rb} 
\frac{\delta\hat{\rho}}{\sqrt{\rho_0}} \Delta\sqrt{\rho_0} \\
H_{\rm kin}^{(2,0)} &=& -\frac{\hbar^2}{2m} l^D \sum_{\rb}
\left[\frac{\delta\hat{\rho}}{4\sqrt{\rho_0}}\Delta\frac{\delta\hat{\rho}}{\sqrt{\rho_0}}
-\frac{\delta\hat{\rho}^2}{4\rho_0^{3/2}}\Delta\sqrt{\rho_0} \right] \\
H_{\rm kin}^{(3,0)} &=&  -\frac{\hbar^2}{2m} l^D \sum_{\rb} \left[
\frac{1}{8}\frac{\delta\hat{\rho}^3}{\rho_0^{5/2}}\Delta\sqrt{\rho_0}
-\frac{1}{8}\frac{\delta\hat{\rho}^2}{\rho_0^{3/2}}\Delta
\frac{\delta\hat{\rho}}{\rho_0^{1/2}}\right].
\end{eqnarray}
The second order term of vanishing order in $\epsilon_1$ is also immediately
obtained:
\begin{equation}
H_{\rm kin}^{(0,2)} = \frac{\hbar^2}{2ml^2} l^D \sum_{\rb,j} \sqrt{\rho_0\rho_{0,+j}}\,
(\hat{\theta}_{+j}-\hat{\theta})^2.
\end{equation}
The last second order quantity to calculate is $H_{\rm kin}^{(1,1)}$, which is
first order in $\epsilon_1$ and first order in $\epsilon_2$.
There are four terms, two involving $\hat{\theta}_{+j}$ and two being their
hermitian conjugate. One can then
collect the terms to form commutators:
\begin{eqnarray}
H_{\rm kin}^{(1,1)} &=& -\frac{\hbar^2}{2ml^2} l^D \sum_{\rb,j}
\frac{i}{2}\left(\frac{\rho_0}{\rho_{0,+j}}\right)^{1/2}
[\hat{\theta}_{+j}-\hat{\theta},\delta\hat{\rho}_{+j}]
-\frac{i}{2}\left(\frac{\rho_{0,+j}}{\rho_0}\right)^{1/2}
[\hat{\theta}_{+j}-\hat{\theta},\delta\hat{\rho}]. \\
&=&  -\frac{\hbar^2}{4ml^2}  \sum_{\rb,j} \left[
\left(\frac{\rho_{0,+j}}{\rho_0} \right)^{1/2}
+\left(\frac{\rho_0}{\rho_{0,+j}}\right)^{1/2}
\right]
\end{eqnarray}
where we have used the commutation relation of $\hat{\rho}$ and $\hat{\theta}$,
see (\ref{commu2}).

We collect all the second order c-number contributions to the Hamiltonian $\bar{H}$
in a single energy functional of the density profile of the quasi-condensate,
\begin{equation}
E_2[\rho_0] =-\frac{g_0}{2}\sum_{\rb} \rho_0-\frac{\hbar^2}{4ml^2} 
\sum_{\rb,j} \left[
\left(\frac{\rho_{0,+j}}{\rho_0} \right)^{1/2}
+\left(\frac{\rho_0}{\rho_{0,+j}}\right)^{1/2}
\right].
\end{equation}

The technique used to calculate $H_{\rm kin}^{(1,1)}$ can be extended to the
calculation of $H_{\rm kin}^{(2,1)}$. There are now three terms and their
hermitian conjugates. Two of these terms, when combined with their hermitian
conjugate, form a commutator that is calculated according to (\ref{commu2}).
The third term and its hermitian conjugate involve the expression
\begin{equation}
\delta\hat{\rho}(\hat{\theta}_+-\hat{\theta})\delta\hat{\rho}_+
-\delta\hat{\rho}_+(\hat{\theta}_+-\hat{\theta})\delta\hat{\rho}
= \delta\hat{\rho}[\hat{\theta}_+,\delta\hat{\rho}_+]- 
[\delta\hat{\rho},\hat{\theta}]\delta\hat{\rho}_+
\end{equation}
which is a sum of two commutators, easy to evaluate. This leads to
\begin{equation}
H_{\rm kin}^{(2,1)} = 
\frac{\hbar^2}{8m} \sum_{\rb}
\frac{\delta\hat{\rho}}{\rho_0}
\left( \rho_0^{-1/2}\Delta\rho_0^{1/2}-\rho_0^{1/2}\Delta\rho_0^{-1/2} \right).
\end{equation}

To calculate $H_{\rm kin}^{(1,2)}$ we first evaluate 
\begin{equation}
H_{\rm kin}^{(\leq+\infty,2)} = \frac{\hbar^2}{4ml^2}\sum_{\rb,j} l^D
\left[\sqrt{\hat{\rho}}(\hat{\theta}_{+j}-\hat{\theta})^2
\sqrt{\hat{\rho}_{+j}} +\mbox{h.c.}
\right]
\end{equation}
and we expand to first order in $\delta\hat{\rho}$, which leads
to a sum of terms that are not individually hermitian. We then use the commutation
relation (\ref{commu2}) to produce hermitian terms, e.g.
\begin{equation}
\delta\hat{\rho}(\hat{\theta}_{+j}-\hat{\theta})^2 =
(\hat{\theta}_{+j}-\hat{\theta}) \delta\hat{\rho} (\hat{\theta}_{+j}-\hat{\theta})
-\frac{i}{l^D}(\hat{\theta}_{+j}-\hat{\theta}).
\end{equation}
The last term of the right hand side of this expression is antihermitian and does
not contribute to the final result
\begin{equation}
H_{\rm kin}^{(1,2)} = \frac{\hbar^2}{4ml^2}
\sum_{\rb,j} l^D\,
(\hat{\theta}_{+j}-\hat{\theta})
\left(\frac{\rho_{0,+j}^{1/2}}{\rho_0^{1/2}}\,\delta\hat{\rho}+
\frac{\rho_0^{1/2}}{\rho_{0,+j}^{1/2}}\,\delta\hat{\rho}_{+j} \right)
(\hat{\theta}_{+j}-\hat{\theta}).
\end{equation}

Finally $H_{\rm kin}^{(0,1)}$ and $H_{\rm kin}^{(0,3)}$ vanish as the odd order expansion of
$\exp[i(\hat{\theta}_{+j}-\hat{\theta})]$ is antihermitian.

\section{Corrections to the equations of motion due to $H_3$}
\label{appen:mvt}

The Hamiltonian $H_3$ gives rise to 
quadratic corrections to the equations of motion for $\delta\hat{\rho}$
and $\hat{\theta}$. In this appendix,  
these corrections are calculated explicitly and the thermal average is taken over 
the equations of motion with the Hamiltonian $H_2 + H_3$ for the linear 
part and the Hamiltonian $H_2$ for the quadratic corrections. This allows
to calculate the first correction to the mean density due to $H_3$.

The corrections to the equation of motion for the density fluctuations
are given by:
\begin{equation}\label{eq:quadracorrec}
\hbar \partial_t\delta\hat{\rho}|_{H_3} =
\frac{\hbar^2}{4ml^2}\sum_j \left[
\left\{\hat{\theta}-\hat{\theta}_{+j},
\left(\frac{\rho_{0,+j}}{\rho_0} \right)^{1/2}\delta\hat{\rho}+
\left(\frac{\rho_0}{\rho_{0,+j}} \right)^{1/2}\delta\hat{\rho}_{+j}
\right\} + \ \ \left(+j\leftrightarrow -j\right)
\right]
\end{equation}
where $\{A,B\}$ stands for the anticommutator $AB+BA$ of two operators. 
When we take the average with the Hamiltonian $H_2$, we use  the explicit modal
expansion of $\dro$ and $\tet$ given by Eq.(\ref{quant}). The operator
$\hat{Q}$ 
disappears since Eq. (\ref{eq:quadracorrec}) involves only differences of 
$\tet$. Terms with $\hat{P}$ also disappear since  
$\langle \hat{P} \rangle_2 = 0$. The expectation value of
the product $\tet(\rb) \dro(\rbp)$, where $\tet$ is written without 
the $\hat{Q}$ and $\dro$ is written without
the $\hat{P}$, is actually purely imaginary:
as $u_s$ and $v_s$ can be chosen to be real, 
$\theta_s^* = -\theta_s$, see Eq.(\ref{def1}). Since $\partial_t
\langle \delta\hat{\rho}\rangle$ is real, all imaginary contributions
to it have to cancel so that  the
correction to the motion of $\langle \delta\hat{\rho}\rangle $
due to $H_3$ finally vanish when we take the thermal average:
\begin{equation}
\hbar \partial_t\langle\delta\hat{\rho}\rangle|_{H_3} = 0
\end{equation}

The corrections to the equation of motion for $\hat{\theta}$
are more involved:
\begin{eqnarray}\label{eq:quadra2}
\nonumber
\hbar \partial_t\hat{\theta}|_{H_3} = 
\frac{1}{2 \sqrt{\rho_0}}  \left( -\frac{\hbar^2}{4 m \rho_0} \dro \Delta 
\left( \frac{\dro}{\sqrt{\rho_0}} \right)  +\frac{3 \hbar^2}{8 m \rho_0^2} 
\dro^2 \Delta \left(\sqrt{\rho_0}\right) - \frac{\hbar^2}{8 m} \Delta 
\left( \frac{\dro^2}{\rho_0^{3/2}} \right) \right. \\ \left.
- \frac{\hbar^2}{2 m l^2} 
\sum_j \left(  \sqrt{\rho_{0,+j}} (\tet_{+j} - \tet)^2 
+ \sqrt{\rho_{0,-j}} (\tet_{-j} - \tet)^2 \right) + \frac{g_0 \sqrt{\rho_0}}{l^D} 
- \frac{\hbar^2}{4 m l^D \sqrt{\rho_0}} \left( \rho_0^{-1/2}\Delta\rho_0^{1/2}-\rho_0^{1/2}
\Delta\rho_0^{-1/2} \right) \right)
\end{eqnarray} 
Fortunately we can use the linear equations of motion 
(\ref{hydro_1},\ref{hydro_2}) to significantly simplify
the above equation of motion.
We rewrite the first term in parentheses of Eq.(\ref{eq:quadra2}) as:
\begin{equation}\label{des1}
 -\frac{\hbar^2}{4 m \rho_0} \dro \Delta 
\left( \frac{\dro}{\sqrt{\roo}} \right)  = -\frac{\dro}{2 \roo}  \left[ (U-\mu+3g_0\rho_0)\left(\frac{\dro}{\sqrt{\roo}}\right) + 2 \sqrt{\roo} \hbar \partial_t \tet \right]
\end{equation}
The second term in parentheses of Eq.(\ref{eq:quadra2}) gives, as $\sqrt{\rho_0}$ solves
the Gross-Pitaevskii equation:
\begin{equation}\label{des2}
\frac{3 \hbar^2}{8 m \rho_0^2} 
\dro^2 \Delta \left(\sqrt{\rho_0}\right) =  \frac{3 \dro^2}{4 \roo^{3/2}} 
( U-\mu+g_0\roo) .
\end{equation}
The sum of Eqs.(\ref{des1}),(\ref{des2}) and the third term 
in parentheses of Eq.(\ref{eq:quadra2}) lead to:
\begin{equation}
\frac{1}{4}\left( -\frac{\hbar^2 \Delta}{2 m}+ U - \mu + 
g_0 \roo \right)\left( \frac{\dro^2}{\roo^{3/2}} \right) 
- g_0 \frac{\dro^2}{\sqrt{\roo}} -\hbar (\partial_t\tet) \frac{\dro}{\sqrt{\roo}}.
\end{equation}
To rewrite the fourth term in parentheses of Eq.(\ref{eq:quadra2}), 
it is convenient to use the following identity:
\begin{eqnarray}\nonumber
\sqrt{\rho_{0,+j}} (\tet_{+j} - \tet)^2 
+ \sqrt{\rho_{0,-j}} (\tet_{-j} - \tet)^2 + 2 \tet \left( \sqrt{\rho_{0,+j}} 
(\tet_{+j} - \tet) + \sqrt{\rho_{0,-j}} (\tet_{-j} - \tet) \right) &=& 
\sqrt{\rho_{0,+j}} (\tet_{+j}^2 - \tet^2) + \sqrt{\rho_{0,-j}} 
(\tet_{-j}^2 - \tet^2) \\
&=& l^2 \left( \Delta \left( \sqrt{\roo} \tet^2  \right) - \tet^2 \Delta \left( \sqrt{\roo}  \right) \right)
\end{eqnarray}
leading to:
\begin{eqnarray}
\sqrt{\rho_{0,+j}} (\tet_{+j} - \tet)^2 
+ \sqrt{\rho_{0,-j}} (\tet_{-j} - \tet)^2 =  l^2 \left( \tet^2 
\Delta\sqrt{\roo} - 2 \tet \Delta \left( \sqrt{\roo} \tet \right)  +  \Delta \left( \sqrt{\roo} \tet^2  \right) \right).
\end{eqnarray}
Using this equality, the Gross-Pitaevskii equation (\ref{fond}) and the 
equation of motion (\ref{hydro_2}), the fourth term in parentheses  of Eq.(\ref{eq:quadra2})
can be written as:
\begin{equation}
- \frac{\hbar^2}{2 m l^2} 
\sum_j \left(  \sqrt{\rho_{0,+j}} (\tet_{+j} - \tet)^2 
+ \sqrt{\rho_{0,-j}} (\tet_{-j} - \tet)^2 \right) = \left( - \frac{\hbar^2}{2 m} \Delta + U - \mu + g_0 \roo \right)\left(\sqrt{\roo} \tet^2\right) - \hbar \tet \frac{\partial_t \dro}{\sqrt{\roo}}.
\end{equation}
The sixth (and last) term in parentheses of Eq.(\ref{eq:quadra2}) 
can also be transformed using the Gross-Pitaevskii equation (\ref{fond}): 
\begin{equation}
- \frac{\hbar^2}{4 m l^D \sqrt{\rho_0}} \left( \rho_0^{-1/2}\Delta\rho_0^{1/2}-\rho_0^{1/2}
\Delta\rho_0^{-1/2} \right) = - \left( -\frac{\hbar^2}{2 m} \Delta + U - \mu + g_0 \roo \right) \left( \frac{1}{2 l^D \sqrt{\roo}} \right).
\end{equation}
This leads finally to a rewriting of the thermal average of Eq. (\ref{eq:quadra2}) as:
\begin{equation}
2 \sqrt{\roo}\hbar \langle\partial_t\delta\hat{\theta}\rangle|_{H_3} = \left( -\frac{\hbar^2}{2 m} 
\Delta + U - \mu + g_0 \roo \right) 
\left( \frac{\langle\dro^2\rangle_2}{4\roo^{3/2}} + \sqrt{\roo} \langle\tet^2\rangle_2  
-\frac{1}{2 l^D \sqrt{\roo}} \right) 
-g_0 \frac{\langle\dro^2\rangle_2 - \roo/l^D}{\sqrt{\roo}}
-\frac{\hbar\partial_t  \langle\tet \dro \rangle_2 }{\sqrt{\roo}} .
\end{equation}
The last term of this expression can be calculated using Eq.(\ref{quant}). 
The harmonic modes do not contribute since the expectation value of products 
of $\hat{b_s}$ and $\hat{b_s}^{\dagger}$ is time independent. We are left with:
\begin{equation}
\partial_t \langle \hat{Q} \hat{P} \rangle _2
= \partial_t \langle \hat{Q(0)} \hat{P} + t \frac{\mu_0'}{\hbar}  
\hat{P}^2 \rangle_2 = \frac{\mu_0'}{\hbar} \langle \hat{P}^2 \rangle_2
\end{equation}
which gives:
\begin{equation}
-\frac{\hbar\partial_t  \langle\tet \dro \rangle_2 }{\sqrt{\roo}} = 2\mu_0'\langle \hat{P}^2 \rangle_2\partial_{N_0}\sqrt{\rho_0}.
\end{equation}
As a conclusion, the quadratic correction to the first equation of motion can be written as
in (\ref{pour_theta}) if one uses the identities:
\begin{eqnarray}\nonumber
 \hbd \hb &=& \frac{\dro^2}{4\roo} + \roo \tet^2  -\frac{1}{2 l^D} \\
\frac{\dro^2}{\sqrt{\roo}} &=& \sqrt{\roo} \left( \hb + \hbd \right)^2 =   \sqrt{\roo} \left(2\hbd \hb + \hb^2 + 
\hat{B}^{\dagger 2} + \frac{1}{l^D}\right).
\end{eqnarray}

\section{The mean value of $\partial_t\hat{\theta}$ vanishes at equilibrium}
\label{pas_de_courant}
As the field degree of freedom $\hat{Q}$, that is the global phase of the field,
is not subject to a restoring force in $H_2$, it is not totally obvious
that the perturbation $H_3$ cannot set it into permanent motion.
We therefore check this point explicitly here.

The first step is to calculate the mean value of $\hat{P}$ to first order
in $H_3$.  We approximate the unnormalized density operator of the gas at thermal equilibrium
to first order in $H_3$ using perturbation theory:
\begin{equation}
\sigma = e^{-\beta (H_2+H_3)}
=e^{-\beta H_2} -\int_0^{\beta}d\tau\, e^{-(\beta-\tau)H_2}H_3 e^{-\tau H_2}+\ldots.
\end{equation}
$\hat{P}$ commutes with $H_2$ and has a vanishing mean value 
in the thermal state
corresponding to $H_2$ so that, to first order in $H_3$:
\begin{equation}
\langle \hat P\rangle_3 = -\langle \beta \hat{P} H_3\rangle_2.
\end{equation}
The Hamiltonian $H_3$ is a polynomial of degree 3 in $\hat{P}$:
\begin{equation}
H_3 = A_0 + A_1 \hat{P} + A_2 \hat{P}^2 + A_3 \hat{P}^3
\end{equation}
where the $A_i$ are still operators with respect to the harmonic oscillator variables
$b_s$.
This leads to
\begin{equation}
\label{pertub_P}
\langle \hat P\rangle_3 = -\beta \left[ \langle A_1\rangle_2 \langle \hat{P}^2\rangle_2
+\langle A_3\rangle_2  \langle \hat{P}^4\rangle_2  \right].
\end{equation}
 From Wick's theorem, $\langle \hat{P}^4\rangle_2 = 3 \langle \hat{P}^2\rangle^2_2$.

In a second step we calculate $\langle d\hat{Q}/dt\rangle$ to first order
in $H_3$:
\begin{eqnarray}
\langle d\hat{Q}/dt\rangle &\simeq& \langle \partial_{\hat P} (H_2+H_3)\rangle_3 \nonumber \\
&\simeq& \mu_0' \langle \hat{P}\rangle_3 + \langle A_1\rangle_2 +3 \langle A_3\rangle_2
\langle \hat{P}^2\rangle_2
\end{eqnarray}
where the terms coming from $ \partial_{\hat P} H_3$ are calculated in the thermal
state for $H_2$ since they are already first order in the perturbation.
 From  the value of $\langle \hat P\rangle_3$ obtained from (\ref{pertub_P})
and from (\ref{equipart}) we obtain the desired result:
\begin{equation}
\langle d\hat{Q}/dt\rangle_3 = 0
\end{equation}
to first order in $H_3$.

\section{An equation for $\{\hat{\alpha},\hat{\Lambda}\}$}\label{appen:cartago}
In this appendix, we derive the partial differential equation
(\ref{elim_alpha}).
We first note that $\hat{B}_n$, being a sum of eigenmodes of the operator
${\cal L}_{\rm GP}$, obeys the differential equation for the evolution
governed by $H_2$:
\begin{equation}
i \hbar \partial_t \begin{pmatrix} \hb_n \\ \hbd_n  \end{pmatrix} =
{\cal L}_{\rm GP}
\begin{pmatrix} \hb_n \\ \hbd_n.  \end{pmatrix}
\end{equation}
We project this equation orthogonally to $\phi_0$ and along $\phi_0$,
so that we get the quantum analog of equations (E9), (E10) of
\cite{cartago}, with the simplification that $\phi_0(\rb)$ is real:
\begin{eqnarray}
\label{for_lambda}
i \hbar \partial_t \begin{pmatrix} \hat{\Lambda} \\ \hat{\Lambda}^\dagger \end{pmatrix}& =&
\begin{pmatrix} {\cal Q} & 0 \\ 0 & {\cal Q} \end{pmatrix}
{\cal L}_{\rm GP} \begin{pmatrix} \hat{\Lambda} \\ \hat{\Lambda}^\dagger \end{pmatrix}
+(\hat{\alpha}+\hat{\alpha}^\dagger)
\begin{pmatrix}
{\cal Q} g_0 \rho_0 \phi_0 \\
- {\cal Q} g_0 \rho_0 \phi_0 \end{pmatrix} \\
\label{pour_alpha}
i\hbar \frac{d\hat{\alpha}}{dt} &=& l^D \sum_{\rb} g_0 \rho_0  \phi_0
\left(\hat{B}_n +\hat{B}_n^\dagger \right)
= l^D \sum_{\rb} g_0 \rho_0  \phi_0
\left(\hat{\Lambda}+\hat{\Lambda}^\dagger\right).
\end{eqnarray}
We have introduced the projection matrix 
\begin{equation}
\label{def_proj}
\langle\rb| {\cal Q}|\rbp\rangle= \delta_{\rb,\rbp} - l^D\phi_0(\rb)\phi_0(\rbp).
\end{equation}
As $\hat{\alpha}$ is antihermitian, the source term in (\ref{for_lambda}) 
vanishes and one can replace $\hat{B}_n$ by $\hat{\Lambda}$ in (\ref{pour_alpha}).

We use these two equations of motion to calculate the first order
time derivative of $A(\rb) \equiv \langle\{ \hat{\alpha},\hat{\Lambda}(\rb)\}\rangle_2$.
We do not give the intermediate result.
As $A$ is real here, we have the property
\begin{equation}
\langle\{ \hat{\alpha},\hat{\Lambda}^\dagger(\rb)\}\rangle_2 =  - 
\langle\{ \hat{\alpha},\hat{\Lambda}(\rb)\}\rangle_2.
\label{Areel}
\end{equation}
As $\hat{\Lambda}$ is orthogonal to $\phi_0$ one has
\begin{equation}
{\cal Q} A = A.
\end{equation}
All this leads to (\ref{elim_alpha}).

\section{Interpretation of $\chi$ in the number conserving Bogoliubov approach}
\label{appen:CD}
We assume here that the gas is a quasi-pure condensate so that $\phi_0$ is
now the condensate wavefunction in the Gross-Pitaevskii approximation.
We then show that $\chi(\rb)/N_0$, where $\chi$ is defined in (\ref{for_chi}),
essentially coincides with the lowest order deviation of the exact condensate
wavefunction from the Gross-Pitaevskii prediction $\phi_0$. This deviation
was calculated in \cite{yvan2}.

We split $\chi$ in a component orthogonal to $\phi_0$ and a component
colinear to $\phi_0$:
\begin{equation}
\chi(\rb) = \gamma \phi_0(\rb) + \chi_{\perp}(\rb).
\end{equation}
The component $\gamma$ has a simple physical interpretation: we sum
(\ref{resultat}) over $\rb$ after multiplication by $l^D$. If we omit
the grand canonical term (absent in the canonical treatment of \cite{yvan2})
we obtain
\begin{equation}
N= N_0 + 2\gamma  + \delta N
\end{equation}
where 
\begin{equation}
\delta N \equiv l^D\sum_{\rb} \langle \hat{\Lambda}^\dagger\hat{\Lambda}\rangle_2
\end{equation}
exactly coincides with the mean number of non-condensed particles
predicted in \cite{yvan2}.
The physical interpretation of $2\gamma$ is then simple:
\begin{equation}
\delta N_0 \equiv 2\gamma
\end{equation}
is the correction to apply to the pure condensate prediction for the number of condensate particles in
order to recover the correct Bogoliubov prediction.
Applying to (\ref{for_chi}) the matrix ${\cal Q}$ (\ref{def_proj}) projecting orthogonally 
to $\phi_0$ we obtain
\begin{equation}
\left[-\frac{\hbar^2}{2m}\Delta+U+g_0\rho_0-\mu\right] \chi_\perp
+2{\cal Q} g_0\rho_0\chi_\perp + {\cal Q}\left(2g_0\rho_0\gamma \phi_0+\frac{1}{2}S
\right)=0.
\end{equation}
We modify slightly the form of the source term $S$, eliminating the anticommutator:
\begin{equation}
\{\hat{\Lambda}^\dagger(\rbp),\hat{\Lambda}(\rb) \}=
2\hat{\Lambda}^\dagger(\rbp)\hat{\Lambda}(\rb)+
\frac{1}{l^D}\langle\rb|{\cal Q}|\rbp\rangle.
\end{equation}
This leads to the system
\begin{equation}
\begin{pmatrix} {\cal Q} & 0 \\ 0 & {\cal Q} \end{pmatrix}
{\cal L}_{\rm GP}
\begin{pmatrix} \chi_\perp \\ \chi_\perp \end{pmatrix}
+\begin{pmatrix}
{\cal Q} S_{\rm eff} \\ - {\cal Q} S_{\rm eff}
\end{pmatrix} =0
\label{cd1}
\end{equation}
with the effective source term
\begin{equation}
\label{cd2}
S_{\rm eff}(\rb) = g_0\rho_0(\rb) \phi_0(\rb) (\delta N_0-1)+
g_0 N_0\phi_0(\rb) \left(2\langle \hat{\Lambda}^\dagger(\rb)\hat{\Lambda}(\rb)\rangle_2
+\langle \hat{\Lambda}^2(\rb)\rangle_2 \right)
-l^D\sum_{\rbp} g_0\rho_0(\rbp)\phi_0(\rbp)\langle
\left(\hat{\Lambda}(\rbp)+\hat{\Lambda}^\dagger(\rbp)\right)\hat{\Lambda}(\rb)
\rangle_2
\end{equation}
where we used the fact that here $\langle\hat{\Lambda}^2\rangle_2 =
\langle\hat{\Lambda}^{\dagger2}\rangle_2$ since the condensate wavefunction
is real.
Equation (\ref{cd1}) is the steady version of Eq.(95) of \cite{yvan2},
which gives $N$ times the correction to the condensate wavefunction,
and the source term (\ref{cd2}) exactly coincides with the one
of Eq.(96) of \cite{yvan2} if one realizes that $N=N_0$, so that $\delta N_0 = -\delta N$,
in the systematic expansion used in \cite{yvan2}.

\section{Corrections to $g_1$ due to the cubic Hamiltonian}
\label{appen:g2H3}

We calculate the corrections to the first order correlation function
due to $H_3$ using the perturbative formula (\ref{pert_form}).
A first remark is that 
\begin{equation}
H_3(\tau)\equiv e^{\tau H_2} H_3 e^{-\tau H_2}
\end{equation}
is still cubic in the operators $\hat{b}_s$ since one has
\begin{eqnarray}
 e^{\tau H_2} \hat{b}_s e^{-\tau H_2} &=& e^{-\tau\epsilon_s} \hat{b}_s \\
 e^{\tau H_2} \hat{b}^\dagger_s e^{-\tau H_2} &=& e^{\tau\epsilon_s} \hat{b}^\dagger_s
\end{eqnarray}
where $\epsilon_s$ is the energy of the Bogoliubov mode $s$.
The second step is to use Wick's theorem to calculate the expectation
values in the thermal state corresponding to Hamiltonian $H_2$. One can derive
the general formulas
\begin{eqnarray}
\langle A_1 A_2 A_3 e^{i\Delta\theta}\rangle_2  &=& 
\left[\langle A_1 A_2 A_3 i\Delta\theta\rangle_2 +
\langle A_1 i\Delta\theta\rangle_2 \langle A_2 i\Delta\theta\rangle_2
\langle A_3 i\Delta\theta\rangle_2 \right] e^{-\langle(\Delta\theta)^2\rangle_2/2} \\
\langle A_1 A_2 A_3 A_4 e^{i\Delta\theta}\rangle_2  &=&
\left\{\langle A_1 A_2 A_3 A_4\rangle_2 \left[1+\frac{1}{2}
\langle(\Delta\theta)^2\rangle_2\right]
-\frac{1}{2}\langle A_1 A_2 A_3 A_4 (\Delta\theta)^2\rangle_2 
\right.
\nonumber \\
&+&
\left.\phantom{\frac{1}{1}}
\langle A_1 i\Delta\theta\rangle_2 \langle A_2 i\Delta\theta\rangle_2
\langle A_3 i\Delta\theta\rangle_2 \langle A_4 i\Delta\theta\rangle_2
\right\} e^{-\langle(\Delta\theta)^2\rangle_2/2} 
\end{eqnarray}
where the $A_i$ are linear in $\delta\hat{\rho}$ and $\hat{\theta}$ and have a vanishing mean value.
A last point is to realize that some of the obtained terms contain
a larger number of factors equal to $\Delta\theta$ than other ones. 
Since $\Delta\theta$ scales as $1/\sqrt{\rho_0 l^D}$, see e.g.\ the
expression of $\hat{\theta}$ in terms of the mode functions $u_s$, $v_s$
in (\ref{def1}), the terms with an excess of $\Delta\theta$ factors
are higher order in the expansion and are therefore negligible.
Note that strictly speaking this argument is correct provided
that each factor $\langle A_i \Delta\theta\rangle_2$ remains bounded whatever
the distance from {\bf 0} to $\rb$. This can be checked indeed to be the case:
from the form of $H_3$ one sees that
$A_i$ is either $\delta\hat{\rho}$ or the phase difference between two
neighboring points of the grid. 
One can therefore use the approximate identities:
\begin{eqnarray}
\langle A_1 A_2 A_3 e^{i\Delta\theta}\rangle_2  &\simeq&
\langle A_1 A_2 A_3 i\Delta\theta\rangle_2
e^{-\langle(\Delta\theta)^2\rangle_2/2} \\
\langle A_1 A_2 A_3 A_4 e^{i\Delta\theta}\rangle_2  &\simeq&
\langle A_1 A_2 A_3 A_4\rangle_2 e^{-\langle(\Delta\theta)^2\rangle_2/2} .
\end{eqnarray}
This immediately leads to the identities (\ref{id1},\ref{id2}).


\begin{thebibliography}{99}
\bibitem{hint} This can be seen by taking the derivative of (\ref{transla}) with respect
to $\alpha$.
\bibitem{continuous}
We note that the same problem of course arises in the continuous version of the theory;
one has to replace in the above reasoning $\hat{\rho}(\rb)$ by the integral of
$\hat{\rho}(\rbp)$ over a finite volume around point $\rb$.
\bibitem{note_ci} We have used the identity $\int_0^x du\, (1-\cos u)/u = \ln x + C - \mathrm{Ci}(x)$
where the cosine-integral function $\mathrm{Ci}(x)$ tends to zero in the limit $x\rightarrow +\infty$.
\bibitem{note_A} The trick is to use the rewriting $-\rho A(k)/k^2=(1+2\bar{v}_k^2)
(n_k+1/2) -1/2$ where $2n_k+1=\mathrm{cotanh}\,(\beta\epsilon_k/2)$ 
and $1+2\bar{v}_k^2=(\mu+\hbar^2 k^2/2m)/\epsilon_k$.
It is then clear that $A(k)$ can be extended to the domain $k<0$ to form an even $C^\infty$ function of $k$
since one can extend $\epsilon_k$ as an odd $C^\infty$ function of $k$.
\end{thebibliography}
\end{document}